# Microcavity phonoritons – a coherent optical-to-microwave interface

A. S. Kuznetsov*,[1] K. Biermann,[1] A. Reynoso,[2,3,4] A. Fainstein,[2,3] and P. V. Santos[1]

[1] *Paul-Drude-Institut für Festkörperelektronik, Leibniz-Institut im Forschungsverbund Berlin e. V., Hausvogteiplatz 5-7, 10117 Berlin, Germany* *

[2] *Centro Atómico Bariloche & Instituto Balseiro (C.N.E.A.) and CONICET, 8400 S.C. de Bariloche, R.N., Argentina*

[3] *Instituto de Nanociencia y Nanotecnología (INN-Bariloche), Consejo Nacional de Investigaciones Científicas y Técnicas (CONICET)-CNEA,8400 Bariloche, Argentina*

[4] *Departamento de Física Aplicada II, Universidad de Sevilla, E-41012 Sevilla, Spain*

Optomechanical systems provide a pathway for the bidirectional optical-to-microwave interconversion in (quantum) networks. We demonstrate the implementation of this functionality and non-adiabatic optomechanical control in a single, $\mu$m-sized potential trap for phonons and exciton-polariton condensates in a structured semiconductor microcavity. The exciton-enhanced optomechanical coupling leads to self-oscillations (phonon lasing) – thus proving reversible photon-to-phonon conversion. We show that these oscillations are a signature of the optomechanical strong coupling signalizing the emergence of elusive phonon-exciton-photon quasiparticles – the phonoritons. We then demonstrate full control of the phonoriton spectrum as well as coherent microwave-to-photon interconversion using electrically generated GHz-vibrations and a resonant optical laser beam. These findings establish the zero-dimensional polariton condensates as a scalable coherent interface between microwave and optical domains with enhanced microwave-to-mechanical and mechanical-to-optical coupling rates.

## Introduction

Coherent interactions between microwave (GHz) phonons and optical (hundreds of THz) photons enable the control of opto-electronic phenomena at the nano- and ps-scale, interconversion of optical and microwave photons for communication between distant qubits [1–3] as well as optical information transfer in on-chip computational devices [4, 5]. One strategy towards efficient interconversion uses optomechanical interactions [6], i.e., correlations between optical and mechanical degrees of freedom. In this setting, optomechanical systems relying on the coupling between high-frequency vibrations (phonons) and solid-state excitations have become relevant for advanced photonic applications, including the emerging fields of quantum communication [7] and control of various quantum states [8–14], e.g., qubits [15].

In general, coherent interactions between photons and phonons require a large coupling energy as well as low phonon ($\Gamma_{\mathrm{M}}$) and photon ($\gamma_{\mathrm{phot}}$) decoherence rates. High-efficiency coherent transduction between the particles presupposes a single-photon cooperativity

$$C_0 = \frac{4 \times g_0^2}{\gamma_{\mathrm{phot}} \times \Gamma_{\mathrm{M}}} \tag{1}$$

exceeding unity, where $g_0$ is the single-photon optomechanical coupling rate. If, in addition, $g_0 > \{\gamma_{\mathrm{phot}}, \Gamma_{\mathrm{M}}\}$, the phonon-photon interaction enters the *optomechanical strong-coupling* (OSC) regime [6], where a novel optomechanical quasiparticle emerges – the *phonoriton* [16]. The above requirements become relaxed for photon populations $N_{\mathrm{phot}} > 1$, for which $C_0$ is enhanced by a factor $N_{\mathrm{phot}}$.

---

* Corresponding author, e-mail: kuznetsov@pdi-berlin.de

Reaching the OSC regime in the solid-state faces several challenges imposed by the huge mismatch between the phonon ($f_{\rm M}$) and the photon ($f_{\rm phot}$) frequencies $f_{\rm M} << f_{\rm phot}$, typically large values of $\Gamma_{\rm M}$ and $\gamma_{\rm phot}$, dissimilar spatial dimensions (wavelengths) of the optical and phonon modes, and low magnitudes of $g_0$. In this context, *polaromechanical* systems – optomechanical setups utilizing strongly coupled excitons and photons (simply, polaritons [17] or MPs) in monolitic microcavities (MCs) – become an attractive option [18]. These systems benefit from the large deformation potential exciton-phonon coupling and the simultaneous confinement of photons and phonons [19, 20], enabling coherent polaromechanics [21, 22] with near-unity single-polariton cooperativity [23].

In this work, we first demonstrate the OSC between Bose-Einstein condensates (BECs) of polaritons and GHz phonons confined in $\mu$m-sized potential traps within a semiconductor MC. This OSC results in *phonoritons*, which are evidenced by optomechanical self-oscillations (SOs) or *phonon lasing*. The SOs can be accounted for by the deformation potential coupling between the BEC pseudo-spin states mediated by the phonons. We then show that phonoritons can also be stimulated and controlled by piezoelectrically generated GHz phonons as well as optically. Thus, we establish these traps as a scalable and bidirectional optical-to-microwave interface. The implications of these milestones for the coherent control down to the quantum regime are discussed.

## Results

The studies were carried out using an (Al,Ga)As MC with intracavity traps [24] for phonons and polaritons (cf. Fig. 1a, further details in *Methods*). These traps confine $\lambda_C = \lambda * n_c = 810$ nm photons ($n_c$ is the average refractive index of the MC spacer) as well acoustic phonons with wavelenghts of $3\lambda$ and $\lambda$. The relevant confined phonons have either longitudinal (LA) or transverse (TA) preferential polarizations with frequencies $f_{\rm LA}^{(3\lambda)} = 7$ GHz and $f_{\rm LA}^{(\lambda)} = 20$ GHz for the LA modes and $f_{\rm TA}^{(i\lambda)} \approx 0.7 \times f_{\rm LA}^{(\eta\lambda)}$ ($\eta = 1, 3$) for the TA ones (see *SM-V-B*). Here, 0.7 is the ratio between the TA and LA sound velocities along the MC z[001]-direction. The above implies that $f_{\rm TA}^{(\lambda)} \approx 2 \times f_{\rm LA}^{(3\lambda)}$. Finally, a ring-shaped piezoelectric bulk acoustic wave resonator fabricated on the top surface electrically injects long-lived ($1/\Gamma_{\rm M} \approx 300$ ns) monochromatic LA phonons with frequency tunable around $f_{\rm LA}^{(3\lambda)} = 7$ GHz [25]. Figure 1(c) displays a spatial photoluminescence (PL) map of the sample. The bright PL spot close to the center of the resonator ring-shaped aperture corresponds to the emission of the trap. Its PL at low optical excitation power ($P_{\rm Exc}$) displays discrete energy spectrum typical of a particle in a box. The transition to the BEC at high $P_{\rm Exc}$ is accompanied by an energy blueshift and a nonlinear increase of the PL intensity, as detailed in *SM-II-B*. Simultaneously, the linewidth reduces to a record-low value $\gamma_{\rm MP} \approx 0.5$ GHz $\ll f_{\rm LA}^{(3\lambda)}$, cf. Fig.1(c), which enables the *non-adiabatic interaction regime*. Dependence of $\gamma_{\rm MP}$ on $P_{\rm Exc}$ is detailed in *SM-III-B*. In the following we consider two $4 \times 4$ $\mu$m$^2$ traps with different polariton excitonic content of 0.05 (labelled $T_1$) and 0.2 (labelled $T_2$).

***Optomechanical self-oscillations in single traps.*** Signatures of the polariton-phonon interaction can be readily identified in spectral PL maps of the trap $T_2$ ground state (GS) recorded in the BEC regime for increasing $P_{\rm Exc}$ (cf. Fig. 1f). The energy axis is referenced to the main emission line (the zero-phonon line, ZPL). For $P_{\rm Exc} < 200$ mW, the map shows, in addition to the ZPL, a second line displaced by $\sim 2.3 \times f_{\rm LA}^{(3\lambda)}$, which evidences the splitting of the trap GS into two components. The GS degeneracy can be lifted in an asymmetric trap (i.e., non-square) by the non-vanishing effective in-plane momentum induced by the confinement via the so-called longitudinal-transverse pseudo-spin splitting [26], described in *SM-V-A*. In the BEC state, the splitting can be amplified by polariton-polariton interactions between unequally populated pseudo-spin states.

Figure 1f reveals two remarkable optomechanical features for $P_{\mathrm{Exc}} > 200$ mW: the locking of the pseudo-spin state at $f_{\mathrm{TA}}^{(\lambda)}$ and the emergence of sidebands separated by multiples of $f_{\mathrm{LA}}^{(3\lambda)}$. The latter are indicated by the blue arrows in an exemplary profile for $P_{\mathrm{Exc}} = 320$ mW in Fig. 1d. These sidebands are attributed to *phonon self-oscillations* (SOs) – the excitation of a coherent mechanical motion by a time-independent polariton drive. The stimulated phonons backact on polaritons by locking their pseudo-spin splitting to $f_{\mathrm{TA}}^{(\lambda)}$. We can estimate the optomechanical coupling rate ($g$) leading to the sideband formation by taking into account the fact that the amplitude of the $n^{th}$ sideband is proportional to $J_n^2(g/f_{\mathrm{LA}}^{(3\lambda)})$, where $J_n$ is the Bessel function of the $n^{th}$ order [22]. We point out that the ratio of the peak intensities of the sideband at $E/hf_{\mathrm{LA}}^{(3\lambda)} = 1$ and the ZPL is $J_1^2/J_0^2 \approx 0.3$. This ratio implies that $g \sim f_{\mathrm{LA}}^{(3\lambda)}$. Thus, $g > \{\Gamma_{\mathrm{M}}, \gamma_{\mathrm{MP}}\}$, which confirms the OSC character of the coupling and gives a lower estimate of $C \approx 10^4$ according to Eq. 1. Therefore, the OSC evidences the formation of a phonon-exciton-photon quasiparticle – the *phonoriton* [16]. Interestingly, phonoritons involving $\lambda$ and $3\lambda$ phonons can appear simultaneously, indicating that more than one phonon mode can enter the OSC regime.

The pseudo-spin locking at $f_{\mathrm{TA}}^{(\lambda)}$ (rather than at $2 \times f_{\mathrm{LA}}^{(3\lambda)}$) is further corroborated by the GS PL from polaritons with a reduced exciton content, as illustrated for the trap $T_1$ in Fig. 1c. The PL from the pseudo-spin state is weaker and sidebands are not observed. The GS splitting remains, nevertheless, locked at $f_{\mathrm{TA}}^{(\lambda)}$ over a wide range of excitation powers (cf. additional data in Fig. SM-6). SOs are ubiquitous in optomechanics [27, 28]. In polariton systems, they have been reported for processes of optoelectronic [29] and optomechanical [30] nature. In contrast to the former, the SOs demonstrated here involve transitions between the GS pseudo-spin states rather than between confined levels with different orbitals and larger energy separation. Unlike the report [22] – in the present case, SOs are of the first-order nature and, more importantly, emerge in a single trap rather than in an array.

The optomechanical couping between the GS pseudo-spin states leading to SOs requires confined phonons with shear strain components, which are intrinsic for TA modes but absent in bulk LA ones propagating along [001] GaAs. However, if the traps are not perfectly square, the lateral confinement imparts a small shear component to the confined LA modes, which is proportional to the trap asymmetry as shown in *SM-V-C*. Furthermore, a first-order deformation potential interaction between phonons and polaritons of the split GS can provide both the interlevel coupling ($g_{0,\uparrow\downarrow}$), corresponding to the coupling between the pseudo-spins, which is required to trigger SOs, as well as the intralevel one ($g_{0,\uparrow\uparrow}$) leading to energy modulation and sideband formation. These coupling rates for the TA and LA confined modes are summarized in Table SM-V for a trap with $a = 4$ $\mu$m and asymmetry $\Delta a/a = 0.1$. In essence, for the GS, the interlevel coupling is considerably higher for TA modes $g_{0,\uparrow\downarrow,\mathrm{TA}} \approx 1$ MHz $\approx 35 \times g_{0,\uparrow\downarrow,\mathrm{LA}}$. Hence, TA-like $f_{\mathrm{TA}}^{(\lambda)}$-phonoritons can form for polariton populations $N_{\mathrm{MP}} \leq 1000$, significantly lower than the BEC threshold of $\sim 10^5 - 10^6$, as estimated in *SM-II-C*. Since $g_{0,\uparrow\uparrow,\mathrm{TA}}$ is negligible for the TA modes, TA-related SOs are normally not accompanied by sidebands. In contrast, LA-like SOs are usually accompanied by sidebands due to the large on-site coupling energy $g_{0,\uparrow\uparrow,\mathrm{LA}} \approx 7$ MHz, but require a large BEC population. These predictions are in qualitative agreement with the results in Figs. 1c and 1d.

Phonoriton formation also requires a pseudo-spin energy splitting $\Delta E$ matching the phonon energy. The following picture emerges for the onset of the phonoriton-related SOs: $\Delta E$ depends on the trap geometry and can change with polariton density to match the phonon energy and trigger a particular phonoriton mode. The matched phonon mode may vary with $P_{\mathrm{Exc}}$ leading to the behaviour illustrated in Fig. 1f. The energy locking between the pseudo-spin states is attributed to the phonon-mediated transfer of particles between them. The strong dependence of the transfer on $\Delta E$ tends to equilibrate the difference in populations leading to the locking [31, 32].

Lastly, SOs can also be induced by interactions between higher-energy (excited) BEC and phonon modes. Some of these phonons can trigger oscillations with just a few polaritons (cf. Table SM-V and the discussion in *SM-V-D*), thus opening the way to SO in the single-particle regime. Furthermore, phonons also affect non-linear polariton interactions [33, 34], including those involving the excitonic reservoir [29, 35]. Combined with the optomechanical coupling proposed here, these mechanisms may additionally enhance the polaromechanical coupling.

***Electrically stimulated sidebands.*** A unique feature of our platform is the ability to electrically inject GHz LA bulk acosutic waves (BAWs) into traps using bulk acoustic wave resonators (BAWRs). Figure 1e shows a spectrum of the trap $T_2$ under the modulation by $f_{\mathrm{LA}}^{(3\lambda)} = 7$ GHz phonons generated by the BAWR driven with radio frequency (RF) voltage. One now observes well-defined and symmetric sidebands separated by $f_{\mathrm{LA}}^{(3\lambda)}$. This demonstrates the non-adiabatic control of the polariton BEC by the tunable phonon amplitude.

The evolution of the PL spectrum of the trap $T_1$ for increasing acoustic amplitudes $A_{BAW}$ is illustrated by the color map of Fig. 2a and cross-sections in Fig. 2b-e. $A_{BAW}$ is expressed in terms of the square-root of the nominal RF power ($P_{\mathrm{RF}}^{0.5}$) applied to the BAWR. Spectra at low acoustic amplitudes $P_{\mathrm{RF}}^{0.5} < 0.01\ \mathrm{W}^{0.5}$ are dominated by the strong ZPL with a weaker pseudo-spin state locked at $f_{\mathrm{TA}}^{(\lambda)} \approx 2 \times f_{\mathrm{LA}}^{(3\lambda)}$ indicated by the red arrow in Fig. 2b. At $P_{\mathrm{RF}}^{0.5} \approx 0.01\ W^{0.5}$, the first two symmetric sidebands appear on either side of the ZPL. At higher acoustic amplitudes, additional symmetric sidebands emerge, reaching up to $\pm 5 \times f_{\mathrm{LA}}^{(3\lambda)}$-sidebands. For the intermediate $P_{\mathrm{RF}}^{0.5}$ values, such as in Fig. 2d, the intensity of the ZPL line becomes strongly suppressed. This suppression is a form of optomechanically induced transparency.

The solid blue lines in Figs. 2b-e are fits given by a sum of Lorentzians with linewidths ($\delta E$) weighted by squared Bessel functions $J_{\mathrm{n}}^2(\chi)$, where $\chi$ – is the modulation amplitude (see *SM-IV-B*). The fits show that the acoustic modulation redistributes the oscillator strength (initially at the ZPL) among the sidebands while conserving the overall PL intensity. Figure 2g shows the dependence of the fitted sideband linewidths $\delta E$ on the normalized $A_{\mathrm{BAW}}$. Remarkably, $\delta E(A_{\mathrm{BAW}})$ sharply decreases by a factor of two from $\delta E(0.1) = 0.2 \times f_{\mathrm{LA}}^{(3\lambda)}$ to $\delta E(0.2) = 0.1 \times f_{\mathrm{LA}}^{(3\lambda)}$ and then remains constant. The reduction coincides with the appearance of the first sidebands, i.e., when $\chi \approx f_{\mathrm{LA}}^{(3\lambda)}$, cf. Fig. 2f. A similar linewidth reduction is also observed for the first excited state of the trap (*SM-IV-C*).

The OSC between particles with largely dissimilar lifetimes leads to quasiparticles with lifetimes approximately twice the one of the shorter-lived component. Such a behavior has been previously reported for quasiparticles resulting from the strong-coupling of excitons and photons [36], photons and phonons [6], as well as phonons and superconducting qubits [37], but not between polaritons and phonons. Stimulated *multimode OSC* has been recently demonstrated for a system of two optical modes coupled to multiple mechanical modes and driven using an external laser [38]. In contrast, in this work, polariton BEC states are intrinsic to the MC. Conceptually, the RF-generated phonon field drives coherent oscillations between the polariton states. In the weak-coupling limit, the states do not swap before decaying with the rate $\gamma_{\rm MP} = 1.4$ GHz. In the OSC limit, one enters the *stimulated phonoriton* regime, where the condensate swaps between the pseudo-spin states at a rate $\gg \gamma_{\rm MP}$. In effect, phonoritons spend half of the time as phonons with $\Gamma_{\rm M} \ll \gamma_{\rm MP}$, thus leading to a decay rate $\gamma_{\rm MP}/2$. The linewidth narrowing thus directly proves phonoriton stimulation.

The above picture can be described using the Hamiltonian (derived in *SM-V-E*) for a phonon mode $\Omega_{\rm M} = 2\pi f_{\rm LA}^{(3\lambda)}$ and two BEC modes $\omega_{\rm i}$ with $\rm i = \{l, u\}$ for the lower energy and the upper energy mode of the spin-split GS, respectively:

$$\hat{H} = \hbar\Omega_M \hat{b}^\dagger \hat{b} + \sum_{i=l,u} \hbar\omega_i \hat{a}_i^\dagger \hat{a}_i + \hat{H}_{\rm int}.$$

The interaction term $\hat{H}_{\rm int} = \hbar G_2 \left(\hat{a}_u^\dagger \hat{a}_l + \hat{a}_l^\dagger \hat{a}_u\right)\left(\hat{b}^\dagger + \hat{b}\right)^2$ couples two BEC states separated by $2 \times \hbar\Omega_{\rm M}$. The $\hat{H}_{\rm int}$ is quadratic in phonon amplitude $\left(\hat{b}^\dagger + \hat{b}\right)^2$, since the pseudo-spin splitting matches twice the energy of the injected phonons, and the coupling strength is tuned by the amplitude of the injected BAW. A detailed analysis of the interaction (cf. *SM-V-E*) yields a coupling strength $g_2 = 2\sqrt{N_{\rm MP} n_{\rm b}} G_2$, where $N_{\rm MP}$ and $n_{\rm b}$ are the polariton and RF-generated phonon populations, respectively, and $G_2 = g_\Delta \times g_{0,\uparrow\downarrow}/f_{\rm LA}^{(3\lambda)}$, where we assumed $g_\Delta \approx g_{0,\uparrow\uparrow}$, see *SM-V-E*. Now, the *OSC condition* becomes $g_2 > \gamma_{\rm MP}/4$. This model predicts a linewidth $\delta{\rm E} = 2 \times {\rm Im}[j \times \gamma_{\rm MP}/4 + \sqrt{g_2^2 - \gamma_{\rm MP}^2}]/\gamma_{\rm MP}$ with $j = \sqrt{-1}$, which is displayed by the solid line in Figure 2g. In the calculations, we used the experimentally determined polariton and phonon decay rates $\gamma_{\rm MP} = 1.4$ GHz, $\Gamma_{\rm M} = 3$ MHz, respectively, and an estimated BEC population $N_{\rm MP} = 5 \times 10^6$. The phonon population $n_{\rm b}$ was determined for each $A_{\rm BAW}$ as described in *SM-II-E*. The calculations reproduce well the linewidth narrowing. However, the fitted $G_2$ value is ∼30 times larger than the one deduced for the pure optomechanical rates in the Table SM-V. The required coupling enhancement can be provided by some of the mechanisms listed at the end of the previous section.

***Coherent optical control of phonoritons.*** Finally, we demonstrate optical control of phonoritons in a trap using the setup depicted in Fig. 3a, which is complementary to the mechanical control addressed in the previous section. For that purpose, a weak single-mode *control laser* with tunable energy $\Delta_L$ was scanned with energy steps of 2.3 GHz across the GS of the trap $T_1$. PL spectra were then recorded for each $\Delta_{\rm L}$ as displayed in Fig. 3b. The weak curved stripes separated by $f_{\rm LA}^{(3\lambda)}$ (indicated by the small yellow arrows) are the phonon sidebands due to the modulation by RF-generated phonons. The weak diagonal feature indicated by the blue dashed arrow is the Rayleigh scattering of the control laser as it was scanned from positive to negative values of $\Delta_{\rm L}$. Interestingly, the sidebands redshift by as much as $0.2 \times f_{\rm LA}^{(3\lambda)}$ when the control laser is within their spectral range, i.e., for $|\Delta_{\rm L}| \leq 5 \times f_{\rm LA}^{(3\lambda)}$. This relatively large red-shift is attributed to a renormalization of the phonoriton GS energy under the increased phonon population induced by the control laser [39].

The most important feature in Fig. 3b is the appearance of $\delta$-like PL peaks whenever the control laser energy matches a sideband, i.e., when $\Delta_L = n_s \times f_{LA}^{(3\lambda)} - \Delta_{rs}$, where $n_s$ is an integer. The latter condition corresponds to the observed enhancement of the integrated emission of the sidebands, cf. Fig. 3(c). Panels (e-h) of Fig. 3 show exemplary profiles recorded for $n_s = \{3, 1, 0, -5\}$. At these laser energies, the amplitude of the sidebands increases up to an order of magnitude as compared to the reference spectrum of Fig. 3d recorded without the control laser, while their linewidth reduces to the resolution limit of 0.5 GHz. The resonant changes of the sidebands around $\Delta_L$ are analogous to the optomechanical heating and cooling, where the scattering of the control laser photons to the sidebands is accompanied by the emission or absorption of phonons. Under RF-generated phonons, however, both scattering processes occur simultaneously, preserving coherence and leading to blue- and red-shifted sidebands with amplitudes exceeding the laser pumped one. We note that the enhancement is selective and affects only some of the sidebands. This behavior is attributed to the interference of two RF-induced sideband combs: one around the phonoriton ZPL and the second one around the control laser energy. The situation is conceptually close to the interference of optical sidebands induced by acoustic waves with different frequencies reported in Ref. [40].

**Discussion and perspectives**

In summary, we demonstrated a compact polaromechanical platform for the coherent conversion between microwave and optical domains. We established that the polariton-phonon interactions are in the OSC regime leading to phonoritons and demonstrated the control of the phonoriton spectrum by electrically generated GHz phonons as well as by an external resonant laser beam. Microscopic models for the optomechanical interaction have also been provided.

The polaromechanical platform opens a new area of GHz phonoritonics. In addition to the optical generation of GHz phonons and the coherent microwave-optical interconversion, we envision that the polaromechanical platform will be attractive for other conventional and emerging applications. Examples include the amplification of optical signals, the generation of tunable (and *symmetric*) optical frequency combs for atomic clocks, high precision spectroscopy, optical synthesizers as well as for the preparation of quantum states [41].

Our results also hint at a far richer and previously unexplored physics, which paves the way to the new interaction regimes between optical, electronic and mechanical degrees of freedom in the solid-state. The large amplitudes of electrically generated GHz strain open the way to phonon nonlinearities, which can be applied for harmonics generation and mixing as well as for parametric processes and phonon squeezing [42]. The observed GS splitting into two pseudo-spin states suggests that the acoustic strain can be used as a source of synthetic magnetic fields for polariton-based topological structures. The results of the work challenge the existing understanding of the polaromechanical interactions, in particular regarding the role of non-linear interactions.

Lastly, a further challenge is to reach single-polariton cooperativities $C_0 \geq 1$ at GHz frequencies by exploiting the large polariton-phonon coupling [18, 43, 44]. For 20 GHz phonons [22, 25], the thermal phonon occupation is $n_{th} \approx 1$ at 1 K, which enables single phonon manipulation at relatively high temperatures. The present structures can already reach $C_0 > 1/20$ (cf. Table SM-V): the current understanding provides pathways to increase $C_0$ by optimizing the trap geometry and material properties. The platform can thus provide coherent control at the single particle level, which can be applied for the generation of non-classical light at GHz rates [45] as well as quantum interfaces between remote polariton qubits [46].

---

## Acknowledgments

ASK and PVS acknowledge the funding from German DFG (grant 359162958) and QuantERA grant Interpol (EU-BMBF (Germany) grant nr. 13N14783). AR and AF acknowledge financial support from the ANPCyT (Argentina) under grant PICT-2018-03255 and the Alexander von Humboldt Foundation.

The authors thank Dr. Stefan Fölsch for discussions and for a critical review of the manuscript as well as the technical support by R. Baumann, S. Rauwerdink, and A. Tahraoui. Data underlying the reported results are included in the main text and supplementary material.

**Supplementary materials**

Supplementary Text

Figs. S1 to S12

Tables I to V

**Methods**

***Microcavity sample.*** In the (Al,Ga)As material system, the sound and light acoustic impedances as well as the ratios between sound and light velocities are almost identical. As a consequence of this "double magic coincidence" [19], an (Al,Ga)As MC designed to confine near-infrared photons also efficiently confines GHz phonons.

Studied MCs consist of the the lower and upper distributed Bragg reflectors (DBRs) and the MC spacer region containing six 15 nm-thick GaAs QWs separated by 7.5 nm-thick $Al_{0.1}Ga_{0.9}As$ barriers. The position of the QWs is optimized in order to maximize the coupling to photonic and phononic modes of the MC. The lower and upper DBRs consist of triple pairs of [(58.1 nm) $Al_{0.1}Ga_{0.9}As$/ (63.1 nm) $Al_{0.5}Ga_{0.5}As$], [(58.1 nm) $Al_{0.1}Ga_{0.9}As$/ (67.6 nm) $Al_{0.9}Ga_{0.1}As$] and [(63.1 nm) $Al_{0.5}Ga_{0.5}As$/ (67.6 nm) $Al_{0.9}Ga_{0.1}As$]. The DBR design provides confinement for optical and acoustic modes with wavelengths $\lambda_o \approx 809/n_{\mathrm{GaAs}}$ nm and $\lambda_a = 3\lambda_o$, respectively. Here, $n_{\mathrm{GaAs}}$ is the GaAs refractive index. The spacer is $3/2\lambda_o$ cavity for photons. The acoustic wavelength corresponds to bulk phonons of $\sim 7$ GHz. Optical and phonon response of the MCs is detailed in *SM-I-B*. The coupling between polaritons and phonons is dominated by the deformation potential interaction.

***Polariton and phonon confinement.*** Structured MCs are fabricated by interrupting the molecular beam epitaxy growth after the deposition of the cavity spacer embedding the QWs and structuring it by photolithographically defined shallow etching [47]. Zero-dimensional confinement regions (the traps) for photons and phonons are defined by nm-high and $\mu$m-wide regions created within the MC spacer [24] (cf. Fig. 1a). We present experimental results recorded on two square polariton traps (traps $T_1$ and $T_2$) with side $a = 4$ $\mu$m and excitonic contents $X^2 = 0.05$ (corresponding to a detuning $\delta_{\mathrm{CX}} = -10$ meV between the bare photon and exciton energies) and $X^2 = 0.2$ ($\delta_{\mathrm{CX}} = -5$ meV), respectively.

***Bulk acoustic wave transducers.*** The phonon generation relies on the transduction of a super-high-frequency (SHF, 3–30 GHz) radio frequency (RF) voltage to sound waves achieved in capacitor-like piezoelectric structures – bulk acoustic wave (BAW) resonators (BAWRs) [48]. A ring-shaped piezoelectric bulk acoustic wave resonator (BAWR) [25] was fabricated on the top surface of the MC to inject monochromatic LA BAWs with frequency $f_M$ tunable around $f_{\mathrm{LA}}^{(3\lambda)} = 7$ GHz into the trap.

An important feature of SHF BAWs is the very weak and essentially frequency-independent acoustic attenuation at temperatures below $\sim 30$ K. This leads to exceptionally long BAW propagation lengths, which reach up to a *cm*. Substrates with polished back-surfaces thus become efficient acoustic cavities with enhanced acoustic amplitudes: here, the BAWs experience specular reflection at the surfaces and make several round trips through the MC spacer before they attenuate. This phonon backfeeding to the MC region boosts the effective quality factor ($Q_{\mathrm{a,eff}}$) to values $Q_{\mathrm{a,eff}} > 10^4$ in the 5–20 GHz range and, hence, to very large $Q_{\mathrm{a,eff}} \times F$ products exceeding $10^{14}$.

***Optical characterization.*** The spatially- and energy-resolved photoluminescence (PL) measurements were carried out at 10 K temperature in a cryogenic (liquid He) cyryostat. The condensates were excited using a single-mode continuous wave (cw) cavity-stabilized laser with the wavelength tuned in the 760–780 nm range. The Gaussian-like excitation spot was chosen to have the diameter of $\sim 40\ \mu m$ on the sample and was centered on the trap. The excitation beam had incidence angle of $\sim 15°$. For the standard measurements with the spectral-resolution of about 0.1 meV, the magnified PL image of the sample was transferred on the entrance slit of a single grating spectrometer and recorded using a Nitrogen-cooled CCD camera.

***High resolution spectroscopy of condensates.*** Figure SM. 3 sketches the high-resolution optical setup used to detect phonon sidebands in the emission of confined condensates. A part of the collected PL was diverted using a mirror and coupled into a single mode (5 $\mu m$ core diameter) fiber. A long-pass filter blocked the scattered light from the pump laser. The fiber guided the PL to a piezo-tunable Fabry-Perot etalon (FP). The FP has a finesse of $\sim 240$ and free spectral range (FSR) of 68 GHz. The transmission wavelength of the FP was tuned by an external voltage source, controlled from a PC. The PL signal filtered by the FP was guided by another single-mode fiber to the entrance of a single grating spectrometer. The latter resolved the PL from the different FSRs of the FP, which was then detected by a nitrogen-cooled CCD. A custom made software was used to control the voltage applied to the FP, which allowed to conduct scans with a resolution of 0.28 GHz. In order to avoid temperature induced drifts, the FP was actively stabilized with an external heater. In this configuration, we loose the spatial information. In order to avoid collecting PL from other traps on the sample, we measured on the sample area, which contained an isolated trap.

***Resonant optical excitation.*** Some experiments were carried out with the simultaneous excitation of the trap with two lasers: the non-resonant pump and a control laser. The wavelength of the single mode (linewidth $\gamma_{\mathrm{L}} \leq 300$ kHz) cw control laser was precisely tuned using a feedback signal provided by a high resolution wavelength meter. The control laser was focused on the same area with the trap into a spot of $\sim 40\ \mu m$ diameter. As schematically shown in Fig. SM 3, the direct reflection of the control laser was blocked.

***Optomechanical coupling in intracavity traps.*** For the determination of the optomechanical couplings, we first calculated the eigenmodes for photons as well as coupled TA and LA phonons confined in a trap with infinite potential barriers and dimensions $a_x = a + \Delta a/2$ and $a_y = a - \Delta a/2$ along the $x||[1\bar{1}0]$ and $y||[110]$ directions, respectively (cf. *SM-V-A* and *SM-V-B*). For that purpose, the photon (phonon) wave functions were expressed in a basis of sinusoidal orbitals with $p_\eta$ ($m_\eta$) lobes along $\eta = \{x, y\}$. The strain field determined from the phonon wave functions was then used to determine the deformation potential coupling to QW excitons using the Pikus and Bir (PB) Hamiltonian [49]. In the last step, we introduced the Rabi coupling between photonic and excitonic modes to determine the polariton eigenstates as well as the optomechanical coupling energies and cooperativities (cf. *SM-V-C* and *SM-V-D*). The quadratic interaction leading to stimulated phonoritons was determined by solving the optomechanical Hamiltonial for two states coupled by the phonons. Expressions for the effective quadratic coupling were then determined in the rotation-wave approximation (cf. *SM-V-E*).

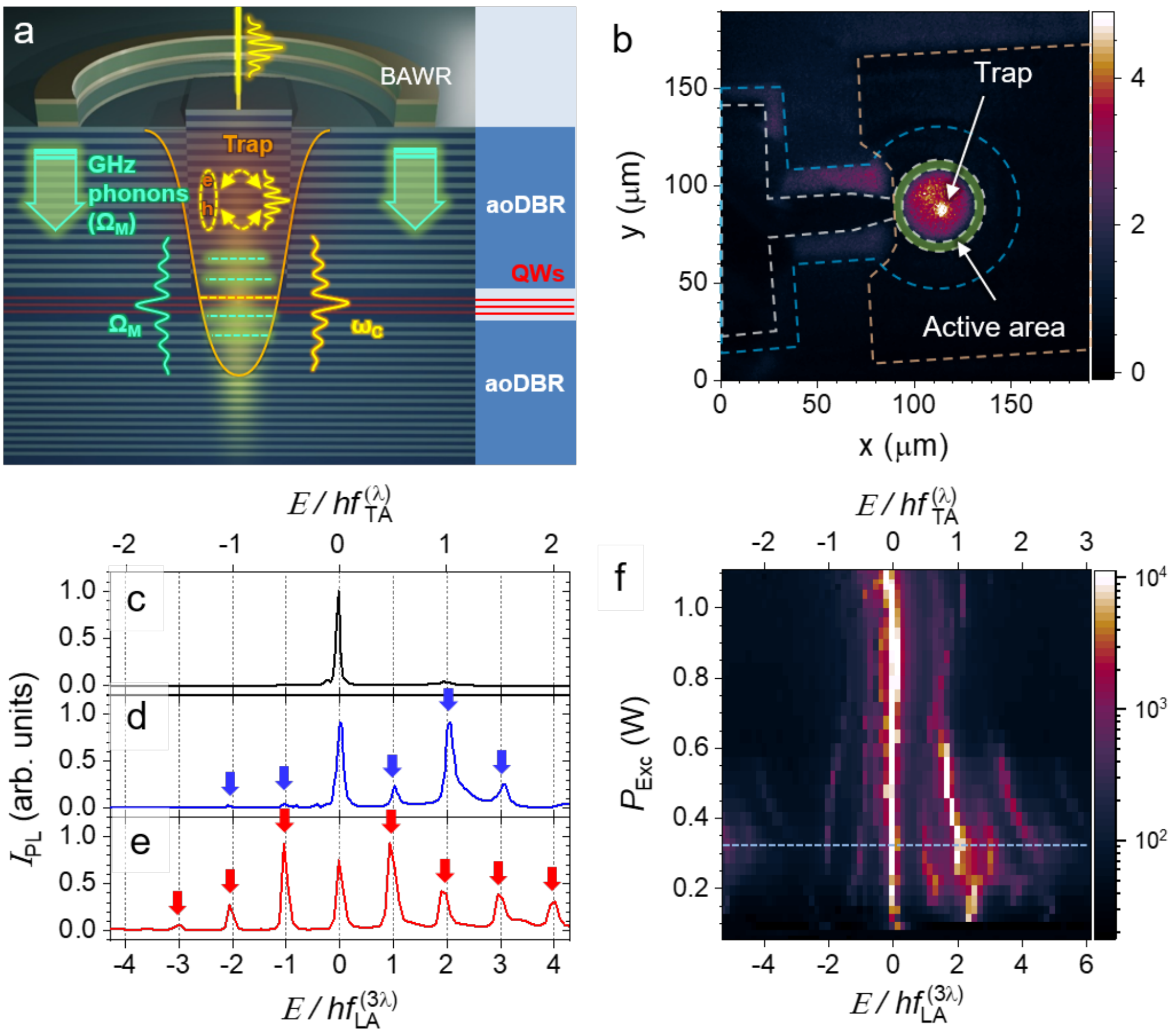

FIG. 1. **Coherent optomechanics with confined polaritons.** **a** Sketch of a structured MC, which consists of a spacer embedding quantum wells (QWs) sandwiched between acousto-optic distributed Bragg reflectors (aoDBRs). The $\mu m$-wide and nm-high mesa within the spacer provides lateral confinement potential (the trap depicted by the yellow curve) for polaritons and phonons. The latter are injected optically or using a ring-shaped piezoelectric bulk acoustic wave resonator (BAWR). The phonons non-adiabatically modulate the discrete polariton energy levels (horizontal dashed yellow line) to form sidebands (dashed green lines). **b** Spatially resolved (energy-integrated) photoluminescence (PL) image map showing the bright emission of trap $T_2$. The superimposed dashed lines are outlines of the BAWR electrodes and its active area. PL spectra of the ground state (GS) of traps **c** $T_1$ and **d** $T_2$ in the BEC regime recorded with the BAWR off. In (d), the self-induced sidebands and indicated by blue-arrows. **e** PL spectrum for $T_2$ with the BAWR driven at 7014.3 MHz displaying induced sidebands (red arrows). **f** Spectral PL map of the GS of $T_2$ for increasing optical excitation and the BAWR off. Curve **d** is a cross-section along the dashed horizontal line. The energy scales in **c-f** are relative to the main PL line and normalized to the phonon energies $hf_{\rm LA}^{(3\lambda)}$ and $hf_{\rm TA}^{(\lambda)}$ in the lower and upper axis, respectively.

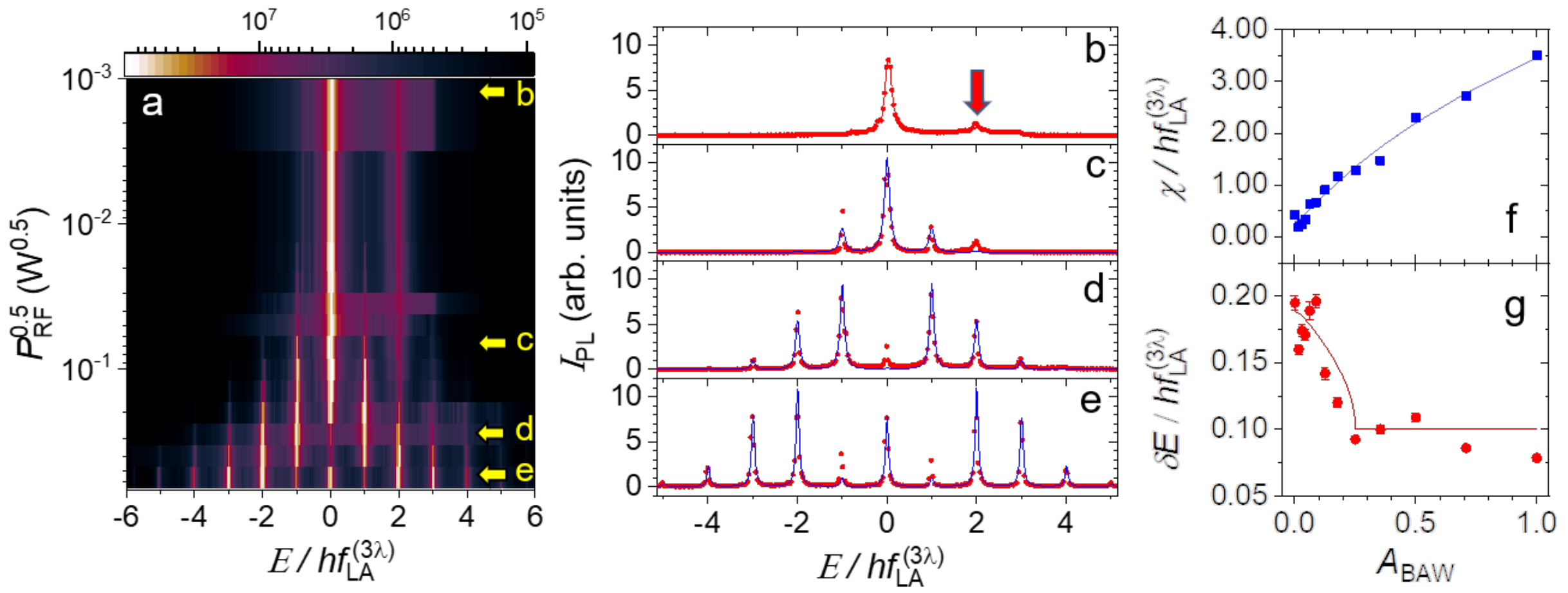

FIG. 2. **Coherent control of a BEC by phonon amplitude.** **a** Dependence of the GS of trap $T_1$ on the acoustic amplitude (proportional to $\sqrt{P_{RF}}$, which is the nominal power applied to the BAWR). **b-e** Cross-sections of **a** along the yellow arrows. The red dots are data. The blue solid lines are fits to the data using the model described in the text. The red arrow mark the pseudo-spin-split state. **f** Dependence of the fitted modulation amplitude and **g** linewidth on the normalized acoustic amplitude, $A_{\text{BAW}}$. The red solid curve is the fitted linewidth as the function of $A_{\text{BAW}}$. The blue solid line in **f** is a guide to the eye.

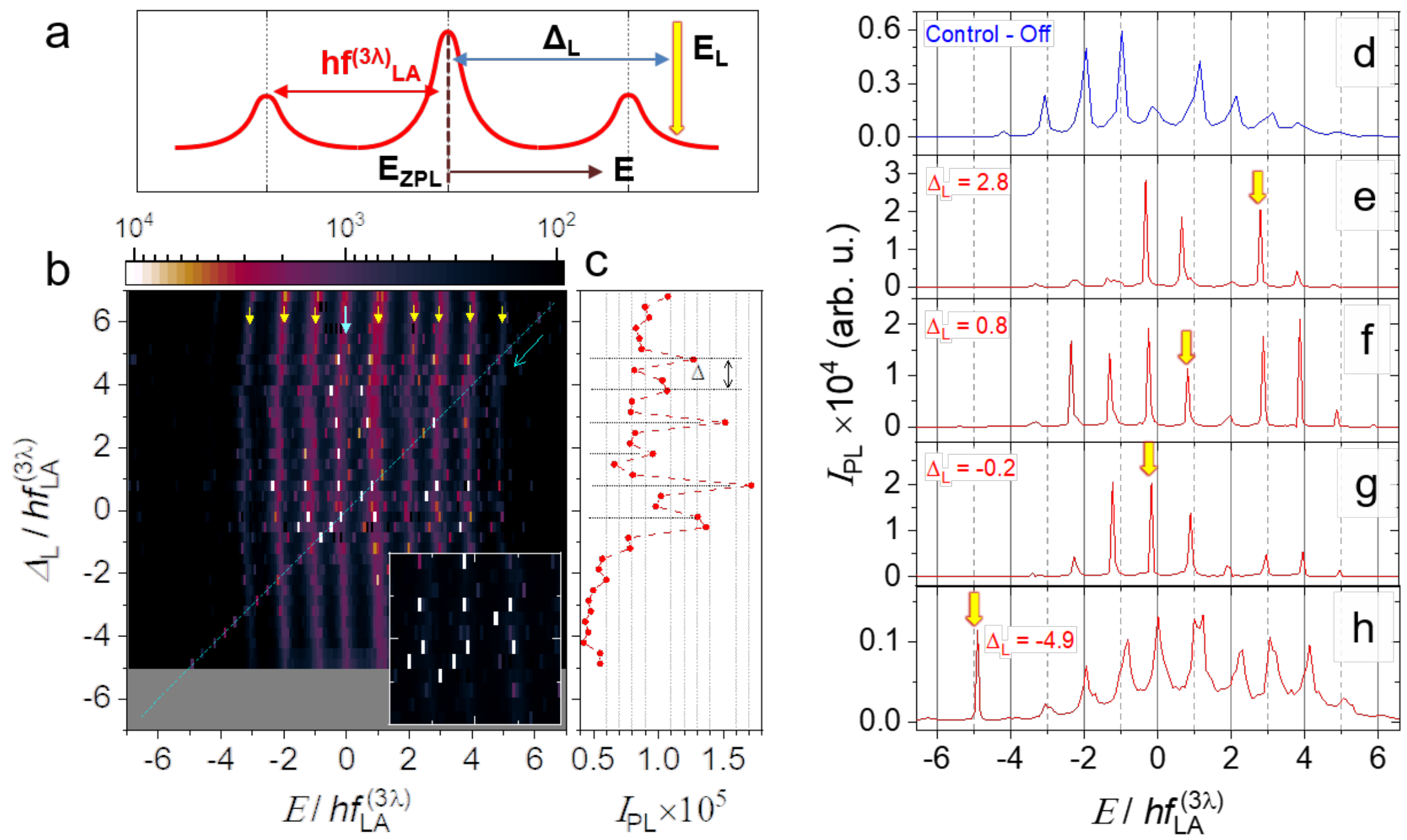

FIG. 3. **Optical coherent control of phonon sidebands.** **a** Experimental setup. **b** Spectral dependence of the GS PL of trap $T_1$ on the control laser energy $\Delta_L$ under excitation of the BAWR. The tiny white rectangles are the $\delta$-like PL peaks induced by the resonant excitation of the sidebands (yellow arrows). The PL energy (horizontal axis) and the control laser detuning (vertical axis) are referenced to the zero-phonon line and scaled to the phonon energy ($hf^{(3\lambda)}_{\mathrm{LA}}$). The solid blue arrow indicates the position of the zero-phonon line. The diagonal line indicated by a dashed blue arrow is the Rayleigh scattering from the control laser. The inset shows a small region of the map plotted with the intensity in linear scale. **c** Integrated PL intensity of the sidebands as a function of $\Delta_L$. **d** Control GS spectrum in the absence of the control laser and **e-h** for selected values of $\Delta_L$ indicated by the yellow arrows.

# Supplementary Material:
# Microcavity phonoritons – a coherent optical-to-microwave interface

A. S. Kuznetsov*,[1] K. Biermann,[1] A. Reynoso,[2,3,4] A. Fainstein,[2,3] and P. V. Santos[1]

[1] *Paul-Drude-Institut für Festkörperelektronik, Leibniz-Institut im Forschungsverbund Berlin e. V., Hausvogteiplatz 5-7, 10117 Berlin, Germany* *

[2] *Centro Atómico Bariloche & Instituto Balseiro (C.N.E.A.) and CONICET, 8400 S.C. de Bariloche, R.N., Argentina*

[3] *Instituto de Nanociencia y Nanotecnología (INN-Bariloche), Consejo Nacional de Investigaciones Científicas y Técnicas (CONICET)-CNEA,8400 Bariloche, Argentina*

[4] *Departamento de Física Aplicada II, Universidad de Sevilla, E-41012 Sevilla, Spain*

## Contents

* Corresponding author, e-mail: kuznetsov@pdi-berlin.de

## I. Polaromechanical microcavities

### A. Detailed structure

The structured polariton microcavity (MC), which is displayed in Fig. 1 of the main text, was grown by molecular beam epitaxy (MBE) on a nominally intrinsic, double-side polished GaAs (001) wafers (Wafer Technology Ltd.). The layer structure was designed to simultaneously confine photons with an optical wavelength $\lambda$ (defined as the ratio $\lambda = \lambda_L/n_s$ between the free space wavelength $\lambda_L = 810$ nm and the effective refractive index, $n_s$, of the MC spacer) and acoustic phonons with wavelengths $\lambda$ and $3\lambda$ [1]. Phonons with these wavelengths will be referred to as $\lambda$-phonons and $3\lambda$-phonons, respectively. For longitudinal acoustic (LA) phonons propagating along the $z||\langle 001\rangle$ direction of GaAs, these wavelengths correspond to phonon frequencies of approx. 21 and 7 GHz, respectively. The layer structure of the sample is summarized in Table I. The spacer region of the MC has an optical thickness $3\lambda/2$ (corresponding to half the LA phonon wavelength) and includes six 15-nm thick GaAs quantum wells (QWs) placed close to the antinodes of the optical field and the acoustic field.

The spacer is sandwiched between a lower (LDBR) and an upper (UDBR) distributed Bragg reflectors (DBRs). Each DBR period consists of a stack of three pairs of $Al_{x_1}Ga_{(1-x_1)}As/Al_{x_2}Ga_{(1-x_2)}As$ layers, each with an optical thickness $\lambda/4$. The thicknesses $d_1$ and $d_2$ and Al compositions $x_1$ and $x_2$ of the layers pairs are listed in Table I. This configuration yields a strong modulation of the optical and acoustic properties with periodicities of $\lambda/2$ and $3\lambda/2$, which enables the simultaneous reflection of both photons with optical wavelength $\lambda$ and phonons with wavelengths $\lambda$ and $3\lambda$.

TABLE I: Layer structure of the microcavity samples. The DBRs consist of stacks of $Al_{x_1}Ga_{1-x_1}As$ and $Al_{x_2}Ga_{1-x_2}As$ layers with thicknesses $d_1$ and $d_2$ and Al compositions $x_1$ and $x_2$, respectively. $d_T$ and $n_{\mathrm{rep}}$ are, respectively, the total thickness and the number of periods.

| Region | $d_T$ (nm) | $n_{\mathrm{rep}}$ | $d_1 : d_2$ (nm) | $x_1 : x_2$ | Comment |
|---|---|---|---|---|---|
| UDBR | 4155 | 11 | 58.12:63.14 | 0.10:0.50 | $\mathrm{pair}_1$ |
| | | | 58.12:67.63 | 0.10:0.90 | $\mathrm{pair}_2$ |
| | | | 63.14:67.63 | 0.50:0.90 | $\mathrm{pair}_3$ |
| growth interruption | — | — | — | — | |
| spacer | 289 | 1 | 289 | 0.10 | including 6 GaAs QWs |
| LDBR | 5289 | 14 | 63.14:67.63 | 0.50:0.90 | $\mathrm{pair}_3$ |
| | | | 58.12:67.63 | 0.10:0.90 | $\mathrm{pair}_2$ |
| | | | 58.12:63.14 | 0.50:0.10 | $\mathrm{pair}_1$ |
| substrate | | | | | |

The intracavity traps are created by interrupting the MBE growth after the deposition of the upper part of the spacer layer (i.e., after the growth of the QWs)[2]. The sample was then extracted from the growth chamber and its surface was photolithographically patterned by shallow (17 nm) etching. It was subsequently reintroduced into the growth chamber for the overgrowth of the UDBR. Due to the conformal nature of the MBE growth, the photolithographically imprinted profile is maintained during the overgrowth, leading to the formation of a thicker spacer region at the mesa sites (denoted here as the non-etched regions) in-between etched areas. The lateral trap confinement results from the lower acoustic and optical cavity resonance energies at these positions. The overgrowth was carried out at a lower deposition temperature (420°C as compared to the temperature of 640°C employed for the first growth cycle) to minize the smoothing of the lateral interfaces due to surface diffusion of impinging atoms.

### B. Optical and acoustic response

The reflection and transmission of optical and acoustic waves through the etched and non-etched regions of the sample were determined using a transfer matrix procedure to calculate the optical and acoustic fields following the excitation (i.e., optical or acoustic) at the sample surface. The results are summarized for the optical and acoustic modes in the left and right panels for Fig. 1.

Figure 1(a) displays the optical reflectivity spectrum $R_{ph}$ for photons impinging at normal incidence on the sample surface. The calculations of the optical response do not include the effects of the excitonic resonances: they yield, therefore, only the bare optical MC mode ($C$). The blue and red curves apply for the etched and non-etched regions of the sample. The high reflectivity band between 1.48 and 1.57 eV is the photon stop-band introduced by the MC DBRs. The sharp dip is the photon cavity mode $C$, which in the etched areas (blue curve) is blue-shifted by 16 meV with respect to the non-etched ones (red curve). The dependence of the optical field for the $C$ mode is illustrated in Fig. 1(d) and in the close-up around the spacer region of Fig. 1(g). Here, the vertical axis displays the squared amplitude of the field $|E(z)|^2$ normalized to the one of the impinging optical beam. The green plots reproduce the depth dependence of the refractive index ($n$) in the MC structure. Here, one can clearly see the $\lambda/2$ and $3\lambda/2$ sequence of DBR layers required for simultaneous photon and phonon reflection. Note that the two sets of three QWs within the spacer are located around antinodes of the optical field, thus ensuring a strong overlap with the QW excitonic resonances.

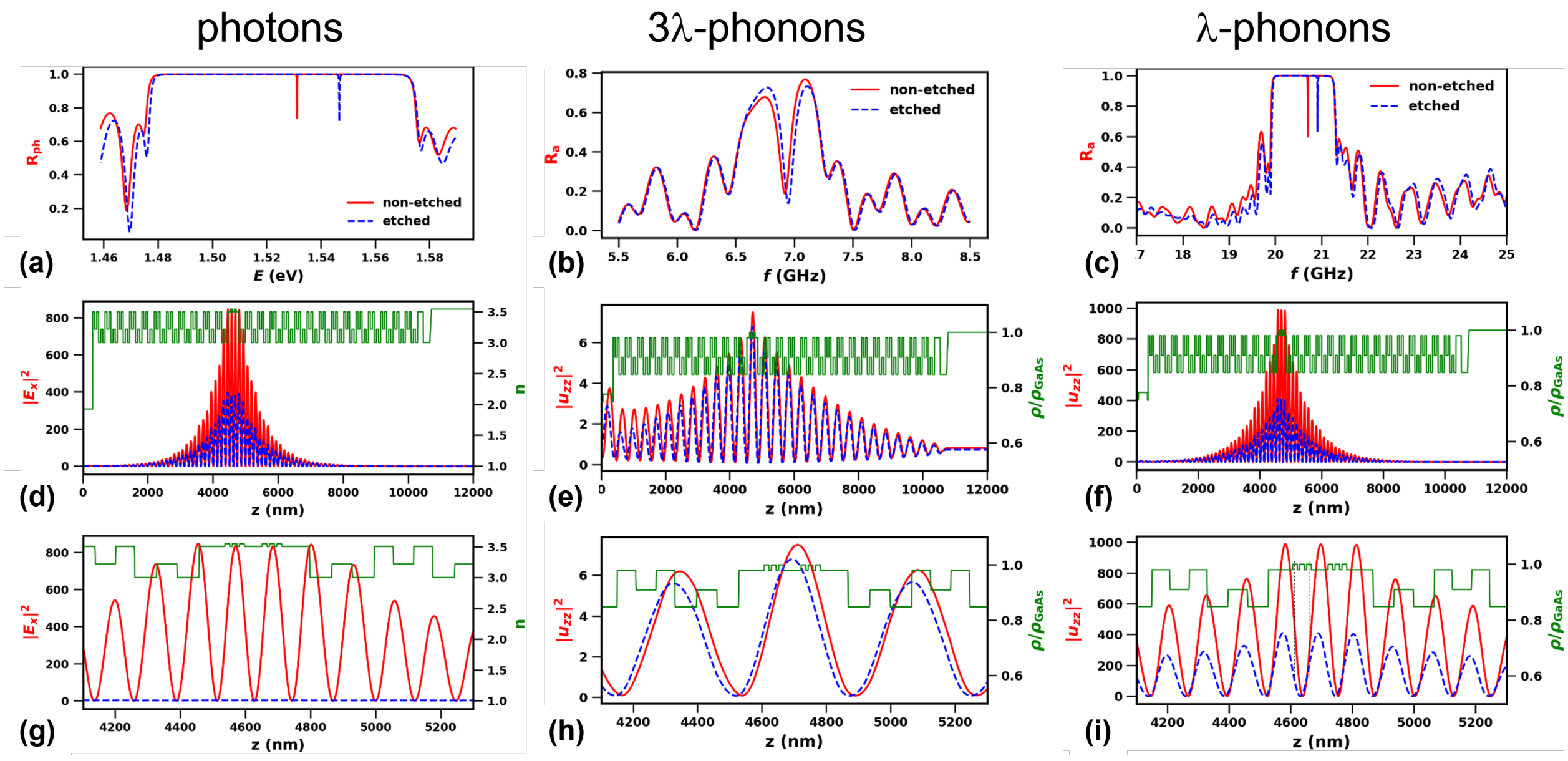

Fig. SM 1: Optical properties of the structured microcavity. (a-c) Calculated optical ($R_{ph}$) reflectivity and acoustic reflectivity ($R_a$) for the longitudinal acoustic (LA) phonons with wavelengths $3\lambda$ and $\lambda$, respectively. The dashed blue and solid red curves are for the etched and non-etched areas of the sample. The dips within the stop-band of high reflectivity between 1.48 and 1.57 eV in (a) 6.6 and 7.3 GHz in (b), and between 20 and 21.5 GHz in (b) and (c) are, respectively, for the polariton, and phonon modes confined within the MC spacer. (d-f) Depth dependence ($z$ direction) of the squared electric $[E^2(z)]$ and strain $(|u_{zz}^2(z)|)$ fields at the reflectivity minima within the stop-bands in (a-c). (g-i) Close-up around the MC spacer area of the profiles in (d-f) within the MC spacer. The green curves in (d-g) display the depth dependence of the refractive index ($n$) of the layers. The green curve in (e-f) and (h-i) displays the corresponding depth dependence for the density $\rho$ (normalized to the GaAs density, $\rho_{\mathrm{GaAs}}$).

The previous picture for photons also applies to phonons. The central and left panels of Fig. 1 show the acoustic reflectivities of the longitudinal acoustic modes with wavelenghts $3\lambda$ and $\lambda$, respectively. The DBRs introduce acoustic stop-bands in the phonon reflectivity ($R_a$) spectrum [Fig. 1(b-c)] extending from 6.6 to 7.3 GHz and between 20 and 21.5 GHz for the $3\lambda$ and $\lambda$ modes, respectively. The dips within the stop-bands are associated with the excitation of modes confined within the MC spacer. The reduced spacer thickness in the nERs blue-shifts the phonon frequency by approximately 27 MHz for the $3\lambda$-phonons and by $\sim$ 100 MHz for the $\lambda$-phonons. The profiles for the squared strain field [Figs. 1(e-f)] are again confined within the spacer: the penetration length into the DBRs is much larger for the $3\lambda$ phonons than for the $\lambda$ ones, where this penetration is comparable to the one for photons in Figs. 1(d). The longer penetration is due to the lower number of effective DBR stacks with $3\lambda/2$ periodicity: as a result, the acoustic quality factor for $3\lambda$ phonons of $Q_a \sim 200$ is much smaller than for photons and for $\lambda$-phonons.

The higher $Q_a$ for the $\lambda$-phonons leads to much narrower resonances in Figs. 1(b-c). Despite the fact that the overlap of the strain field and the QWs has not been optimized for $\lambda$ phonons, their average squared strain field over the QWs exceeds by approximately two orders of magnitude the one for the $3\lambda$-modes cf. Figs. 1(e-f)]. The stronger coupling to excitons resulting from the higher amplitudes of the $\lambda$ phonons has important consequences for self-oscillation effects. As will be discussed later, the phonon backreflection at the sample surfaces considerably increases the quality factors for both types of phonons[1].

## II. Photoluminescence of confined polaritons

### A. Light-matter coupling

Variations of the MBE fluxes along the sample surface create a slight reduction of the thickness of the MBE layers as one move from the center to the border of the substrate wafer. The thickness reduction blue-shifts the optical mode of the MC, while the excitonic resonances remain approximately constant. This relative variation between the optical and excitonic energies enables a precise determination of the Rabi (light-matter) coupling at different positions on the sample surface.

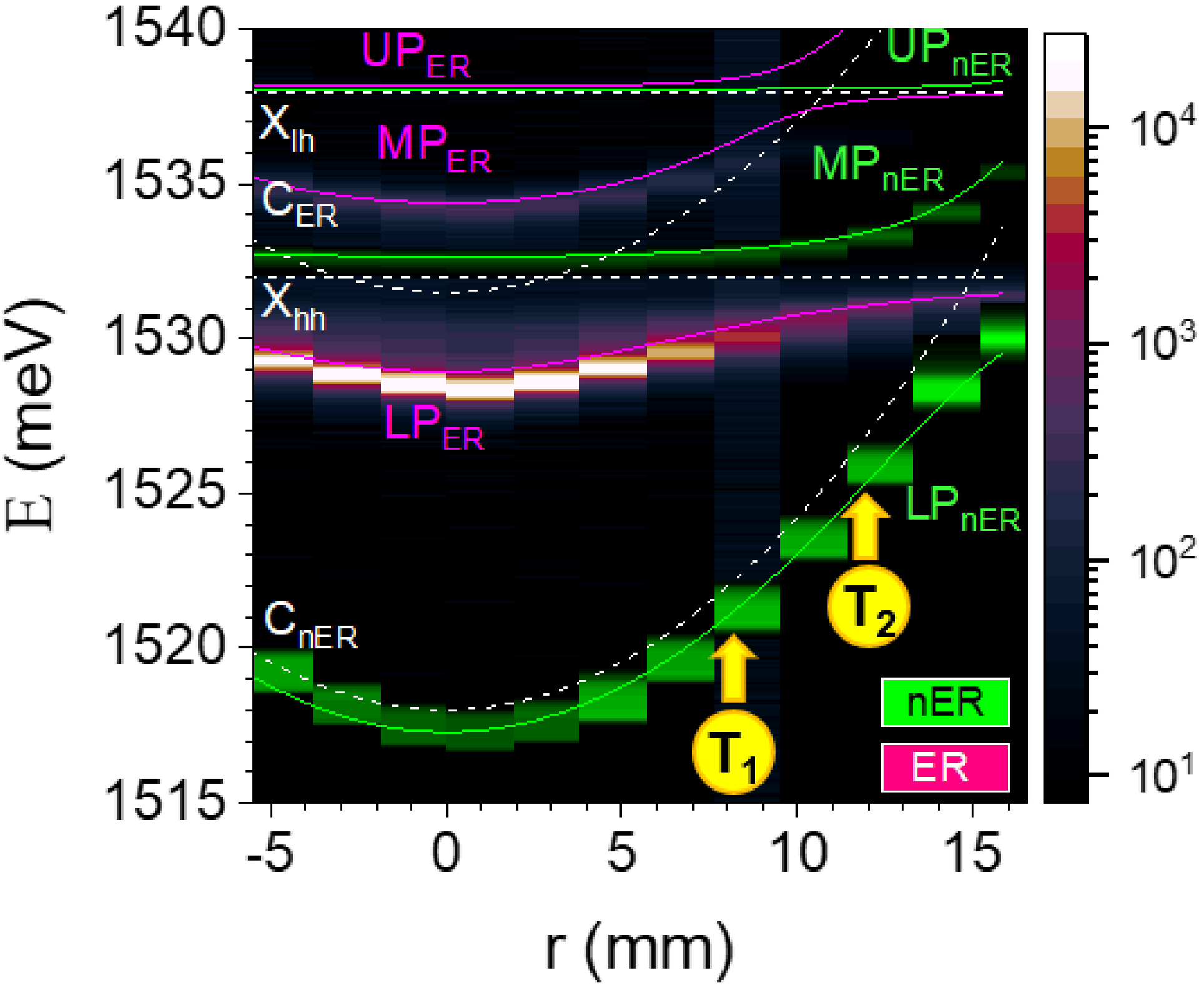

Fig. SM 2: Photoluminescence lines recorded on non-etched (nER, green) and etched (ER, magenta) regions on the 2-inch MC wafer at different radial positions ($r$) relative to the wafer center ($r = 0$). The dashed lines show a fit to a photon-exciton coupled oscillator model for the coupling between the optical mode as well as the hh and lh excitons, yielding the three polariton modes LP (lower polariton), MP (middle polariton), and UP (upper polariton). The dashed lines show the bare photon, hh and lh exciton energies. Note that the former is blue-shifted in the ER regions. $T_1$ and $T_2$ mark the position of the intracavity traps discussed in the main text.

Figure 2 displays the energy of the polariton photoluminescence (PL) recorded on non-etched (green) and etched (magenta) regions as the function of the radial position on the sample (with r = 0 being the wafer center). At each position, one detects PL from the lower (UP), middle (MP), and upper (UP) polariton branches resulting from the coupling between the optical MC mode ($C$) with the electron heavy-hole ($X_{hh}$) and electron light-hole ($X_{hh}$) excitons in the QWs.

The solid lines are fits to a photon-exciton coupled oscillator model for the interaction between the three modes. The model assumes that the reduced spacer thickness blue-shifts the optical modes in the etched regions relative to the non-etched ones, while the excitonic energies remain constant. The fits yields three polariton modes LP, MP, and UP. The parameters used for the fits are summarized in Table II. The dashed lines show the fitted bare $X_{hh}$ and $X_{lh}$ exciton energies, as well as the radial-dependence of the photon energy ($C$) in the non-etched regions. LP is highly photonic at the center of the wafer ($r = 0$). The zero detuning between the $C$ and $X_{hh}$ energies is reached at $r = 15$ mm. Note that the lateral confinement potential for the intracavity traps, which is equal to the difference between the UP energies in the etched and non-etched region, reduces with $r$. $T_1$ and $T_2$ mark the position of the intracavity traps investigated here.

TABLE II: Parameters determined from fits of the spatial dispersion of the polaritons in extended etched (ER) and non-etched (nER) cavity regions.

| Parameter | Value | Comment |
|---|---|---|
| $E_{hh}$ | 1532 meV | Bare heavy-hole exciton energy |
| $E_{lh}$ | 1538 meV | Bare light-hole exciton energy |
| $C_{nER}$ | 1518 meV | Bare cavity mode energy in non-etched region at $r = 0$ mm |
| $C_{ER}$ | 1531.5 meV | Bare cavity mode energy in etched region at $r = 0$ mm |
| $\Omega_{X_{hh}}$ | $6.0 \pm 0.25$ meV | Rabi-splitting energy for cavity-heavy-hole (for both regions) |
| $\Omega_{X_{lh}}$ | $2.0 \pm 0.25$ meV | Rabi-splitting energy for cavity-light-hole (for both regions) |
| $m_{eff}$ | $4.8 \times 10^{-5}$ | Polariton effective mass in free electron mass |

### B. Polariton confinement and condensation

Figure 3a shows an exemplary PL map recorded along the spatial [-1-10] axis of a $4 \times 4$ $\mu m^2$ trap $T_1$. The measurement were carried out at 10 K under non-resonant optical excitation below the condensation threshold. Several confined levels can be easily identified. The spatially integrated spectrum is shown in the middle-section. Each level has a characteristic spatial profile of intensity, which reflects the squared profiles of the polariton wavefunctions. The spectrum can be faithfully reproduced using numerical simulations [2]. The simulated 2D spatial profiles are shown in the right-section.

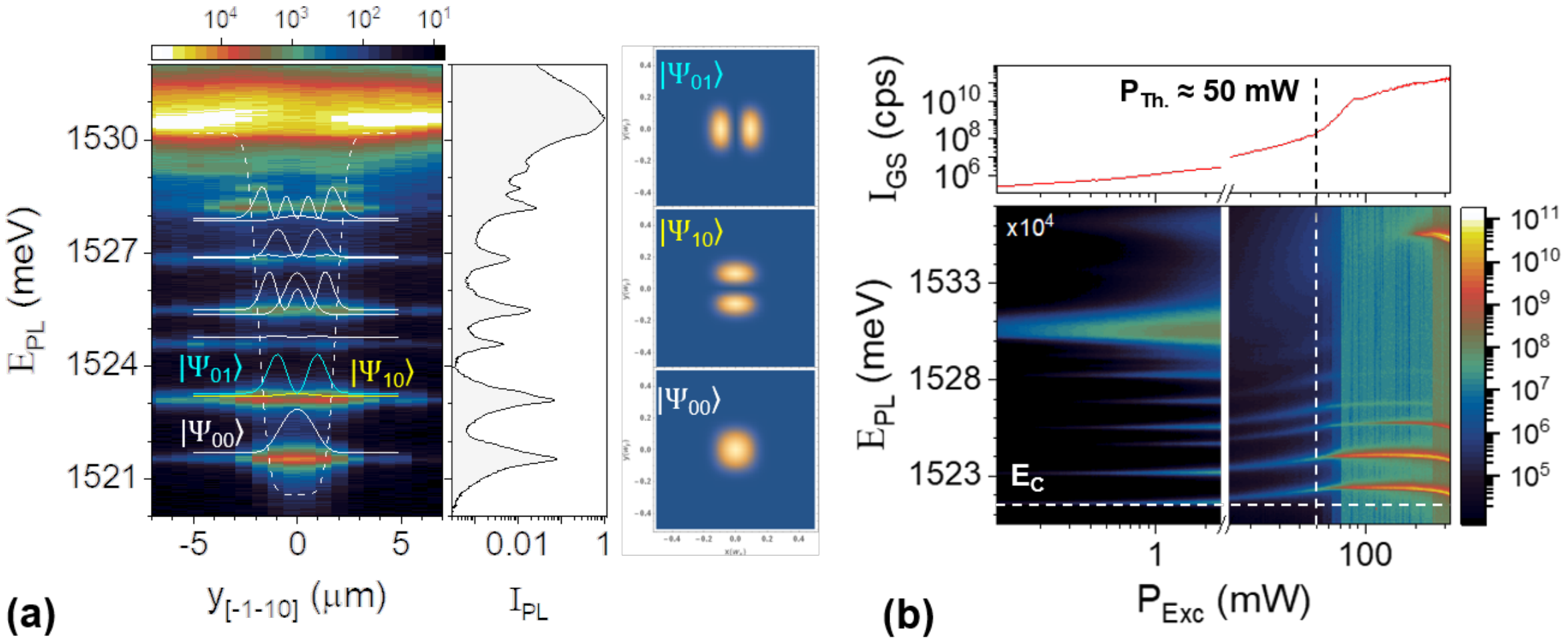

Fig. SM 3: (a) The map of PL of a $4 \times 4\mu$ $m^2$ trap resolved in energy and in space measured at 10K. Several confined polariton modes are visible below the barrier energy around 1530 meV. The dashed and solid white lines on top of the data are the simulated cross-section of the polariton confinement potential and squared wavefunctions, respectively. The section on the right shows a spatially integrated PL spectrum of the trap. The three sections further to the right display calculated spatial profiles of the square wavefunctions for the three lowest energy levels. (b) Lower part: a map showing trap PL spectrum as the function of the optical excitation power ($P_{Exc.}$). Upper part: total integrated PL as the function of the excitation power. The vertical dashed line designates the condensation threshold power ($P_{\mathrm{Th.}}$).

Figure 3b presents the dependence of the trap spectrum on the excitation power of the non-resonant laser. The upper-section shows the integrated intensity of the ground-state (GS). Around $\sim 50$ mW power (dashed vertical line), polaritons transition to the polariton Bose-Einstein condensate (BEC). In the color map, the dashed horizontal line (labelled $E_C$) denotes the energy of the bare cavity mode that was used to fit the spatial spectrum of the trap. The energy difference between the $E_C$ and the condensate in the GS indicates that above the condensation threshold the system is still in the strong-coupling regime.

### C. Estimation of the polariton population

We estimate the optically excited polariton population in a trap as $N_{MP} = (1-R) r_A \eta \frac{P_{Exc}}{\hbar\omega_{Exc}} \tau$. In the latter expression, R is the MC reflectivity at the excitation wavelength, $r_A$ is the ratio between the trap and the excitation spot area, $\eta$ is the fraction of excited electron-hole pairs that are stimulated into the BEC from the exciton reservoir, $P_{Exc}$ and $\hbar\omega_{Exc}$ are the excitation power and photon energy, and $\tau$ is the exciton reservoir lifetime, which is estimated to be on the order of $\tau = 1$ ns. Assuming R = 0.5 for the excitation wavelength of 760 nm, $r_A = 0.01$ for a $4 \times 4$ $\mu m^2$ trap and a Gaussian excitation spot with the radius 20 $\mu m$, and $\eta = 0.1$, we obtain $N_{MP} = 2400 \times P_{Exc}$ for $P_{Exc}$ in $mW$. Hence, at the condensation threshold $P_{Exc} = 50$ mW (cf. Fig. 3), we deduce $N_{MP} \approx 10^5$ for the BEC population.

### D. Acoustic frequency comb in PL of confined polaritons

Figure 4 shows the dependence of the PL spectrum of a trap below the condensation threshold on the radio frequency ($F_{RF}$) applied to the transducer. The spectra were recorded for a fixed RF power. The displayed $F_{RF}$ range corresponds to the range of the $3\lambda$ acoustic mode of the MC. The data shows pronounced energy-modulation of all confined levels, which appears as a comb of narrow diamond-like shapes. Phonons escaping from the spacer region can be re-fed to this region via acoustic reflection at the polished sample surfaces. As a consequence, the acoustic response of the MC is given by a combination of the relatively low quality ($Q_{MC} \approx 180$) MC mode and the higher quality ($Q_{comb} > 5000$) Fabry-Perot modes of the cavity formed by the whole sample thickness [1]. The modulation amplitude increases for the higher confined levels, which have a higher polariton excitonic content: this behavior is consistent with a modulation mechanism dominated by the deformation potential interaction. The envelope of the modulation amplitude gives the shape and the width of the acoustic mode of the MC (localized within the spacer).

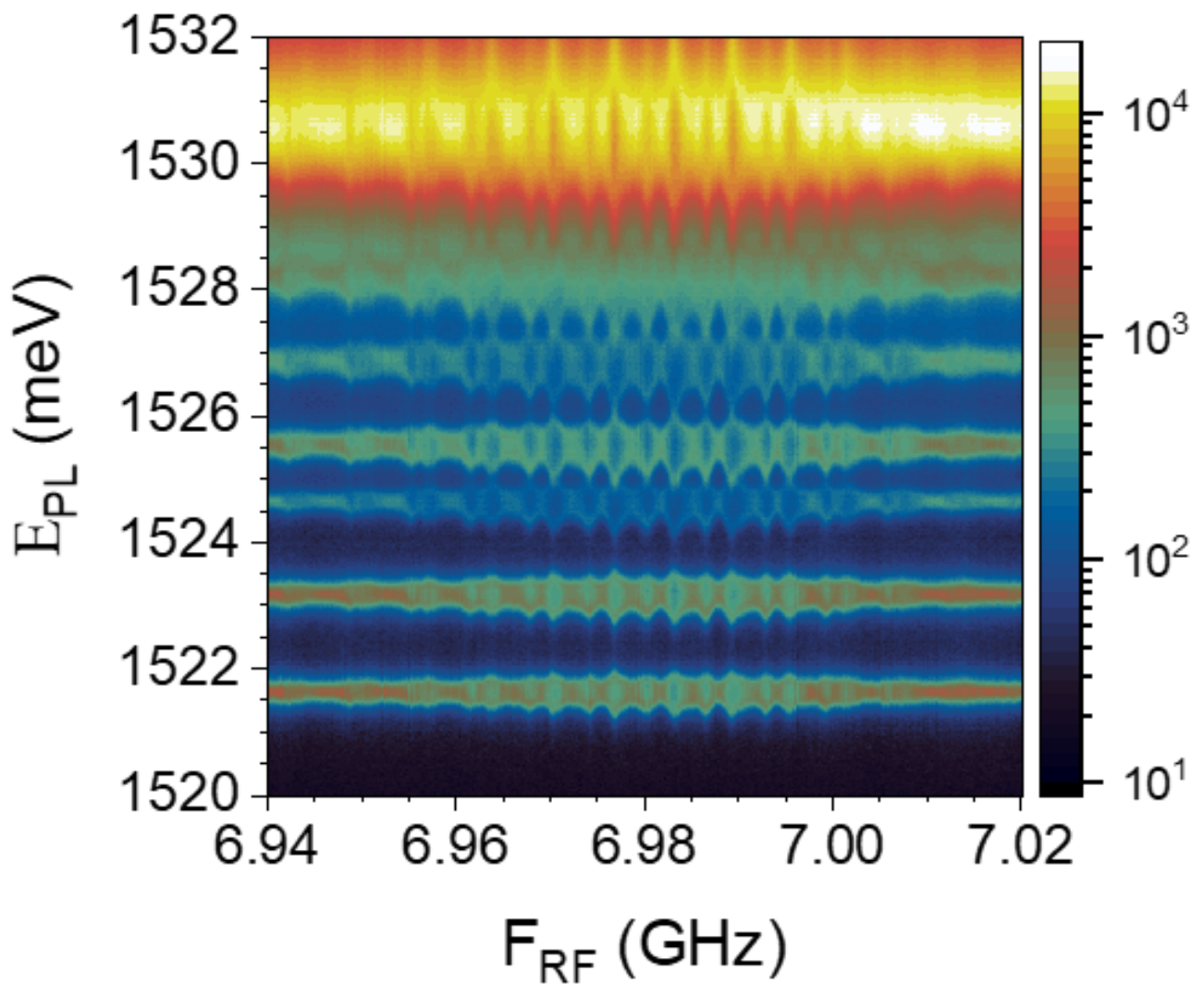

Fig. SM 4: Acoustic frequency comb in PL of confined polaritons. Dependence of the PL spectrum (below the condensation threshold) of a $4 \times 4$ $\mu m^2$ trap on the RF frequency ($F_{RF}$) applied to a BAWR transducer. The $F_{RF}$ range corresponds to the range of the MC acoustic mode.

### E. Population of rf-generated phonons

First, we determine the rf-induced strain from the high-resolution PL spectrum. Specifically, Fig. 2(a) of the main text shows $\pm 4$ phonon sidebands for an rf amplitude of $P_{RF}^{0.5} = 0.14$ $W^{0.5}$. This corresponds to the modulation amplitude of $\Delta E = 4 \times \hbar\Omega_M \approx 114$ $\mu$eV for the phonon frequency of $\Omega_M = 2\pi \times 7$ GHz. The value of the zz-component of the strain corresponding to $\Delta E$ can be found using $\epsilon_{zz} \approx \Delta E/(X^2 \times a_h)$ (cf. Eq. 19 and neglecting the small contribution associated with the deformation potential. For the lower polariton Hopfield coefficient $X^2 = 0.08$ and GaAs hydrostatic deformation potential $a_h$ from cf. Table III, we obtain $\epsilon_{zz} = 1.6 \times 10^{-4}$.

The phonon population in a $4 \times 4 \mu$ $m^2$ trap can be determined as $N_{phon} = \epsilon_{zz}/u_{zz}^{ZPM}$, where $u_{zz}^{ZPM} \epsilon_0 = 1 \times 10^{-8}$ – is the strain of a single 7 GHz phonon, determined from a numerical simulation described in SM Sec. V B (cf. Fig. 12). Thus, we estimate phonon population $N_{phon} = 16000$ for $P_{RF}^{0.5} = 0.14$ $W^{0.5}$.

## III. High-resolution spectroscopy of polariton condensates

### A. Optical setup

Figure 5 sketches a high-resolution optical setup used to detect phonon sidebands in the emission of polariton BECs. The sample is mounted in an optical Helium-flow cryostat with rf-connections for the excitation of the BAWRs. The pump laser and control laser beams are used to excite the sample. A part of the collected PL is diverted using a mirror and coupled into a single mode (5 $\mu m$ core diameter) fiber. A long-pass filter blocks the scattered light from the pump laser. The fiber guides the PL to a piezo-tunable Fabry-Perot etalon (FP). The FP has a finesse of $\sim 240$ and free spectral range (FSR) of 68 GHz. The transmission wavelength of the FP is tuned by an external voltage ($V_{piezo}$) source, controlled from a PC. The PL signal filtered by the FP is guided by another single-mode fiber to the entrance of a single grating spectrometer. The spectrometer resolves the PL from the different FSRs of the FP, which is then detected by a nitrogen-cooled CCD. A custom-made software was used to control the voltage applied to the FP, which allowed to conduct scans with a resolution down to 0.28 GHz. In order to avoid temperature induced drifts, the FP was actively stabilized with an external heater. In this configuration, we loose the spatial information. In order to avoid collecting PL from other traps on the sample, we carried out our measurements on the sample area containing an isolated trap.

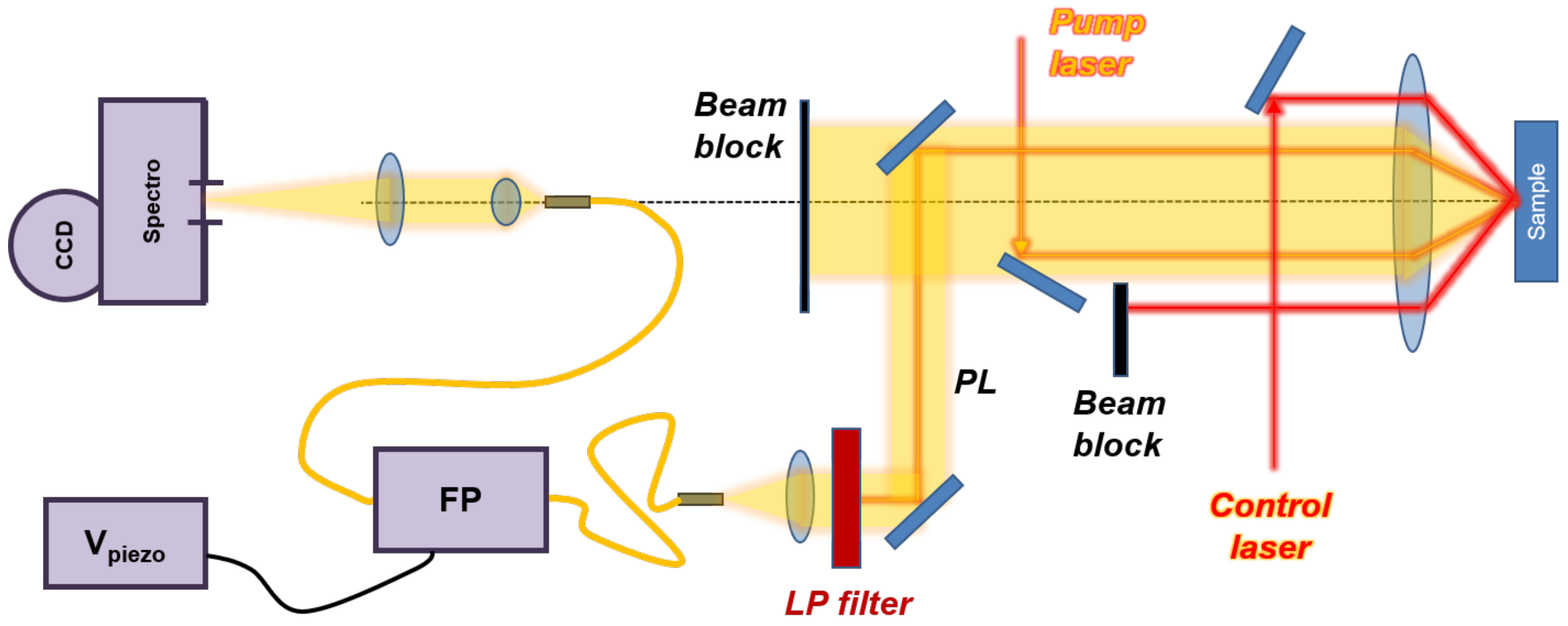

Fig. SM 5: A sketch of the experimental setup for high-resolution optical spectroscopy of phonon sidebands in the emission of confined polariton condensates. The sample is mounted in a liquid He cryostat with rf connection to drive the BAWRs. The orange and red arrows show the optical paths of the pump and control laser beams, respectively, used to excite the sample. The PL signal (represented by the yellow area) is fiber-coupled to a Fabry-Perot etalon (FP) tunable by a piezo-controller ($V_{\rm piezo}$). The output of the FP is directed via a second fiber to a spectrometer with a CCD detector.

### B. Polariton BEC linewidth

Figure 6 displays a high-resolution PL spectrum of the trap ground state (GS) as the function of the pump laser power above the condensation threshold. The trap GS is split into two levels (denoted $L$ and $U$), separated by $2 \times \hbar\Omega_M$, where $\hbar\Omega_M$ is the phonon energy with $\Omega_M/2\pi = 7$ GHz. While their splitting remains constant, $L$ and $U$ modes redshift above $\sim 120$ mW due to the optically-induced heating at high laser excitation. The upper-section of the Fig. 6 shows the decoherence rate ($\gamma_{GS}$) of the resonances determined from their spectral linewidths. The black triangles were determined from a low-resolution measurement (without the FP etalon). These measurements yield $\gamma_{GS} \sim 50$ GHz below the threshold (of $P_{\rm Th.} \approx 50$ mW, as indicated by the orange arrow), which reduces to the resolution limit of $\sim 20$ GHz above the threshold. The blue diamonds and red circles give the corresponding $\gamma_{GS}s$ of the $L$ and $U$ modes obtained with the high-resolution technique based on the piezo-tunable FP etalon. In the range 90–200 mW the $L$ mode has linewidth of $\gamma_L \sim 1$ GHz, which corresponds to a coherence time of $\tau_L \sim 1$ ns. Around 250 mW, the linewidth reaches its minimum $\gamma_L \sim 0.5$ GHz (2 ns). The increase of the linewidth for excitation above 250 mW is attributed to heating.

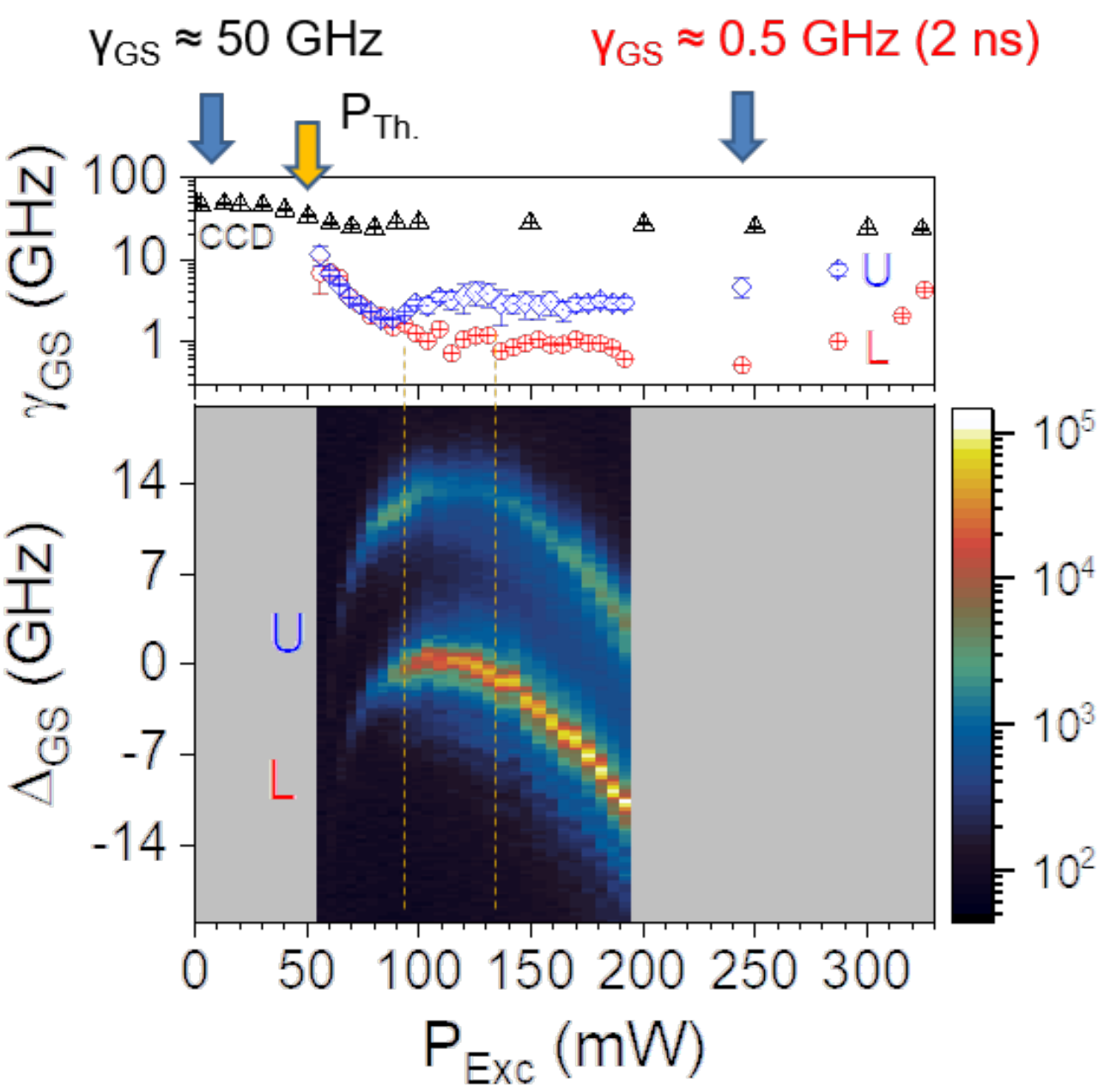

Fig. SM 6: (Lower panel) High-resolution spectrum of the ground state (GS) of the trap as the function of the optical excitation power ($P_{Exc}$). The lower and upper split levels are designated $L$ and $U$, respectively. (Upper panel) GS linewidth ($\gamma_{GS}$) measured with low-resolution of $\sim$ 25 GHz (black triangles), and with high-resolution of $\sim$ 0.3 GHz for the $L$ and $U$ states – red circles and blue diamonds, respectively.

### C. Self-induced sideband modulation

The color coded plot of Fig. 7(a) displays as-measured high-resolution spectra of the condensate emission of the $4 \times 4$ $\mu m^2$ trap as a function of the optical excitation power, $P_{\rm exc}$. These high-resolution measurements were recorded using the FP setup described above. The energy scale is specified in terms of the voltage applied to the piezo controlling the etalon. The figure shows a series of sidebands around the main emission line – the zero-phonon line (ZPL, marked by a red arrow). The spectroscopic features repeat with a periodicity determined by the free spectral range (FSR) of the etalon. All lines experience a small blueshift (less that the FSR) with increasing $P_{\rm exc}$, which is attributed to polariton-polariton interactions. Figure 7(b) displays the same data after artificially aligning the ZPL of all the curves.

The same measurement also captures the emission of the excited state (ES) of the trap, as illustrated in Fig. 7(c). For this particular detuning, the condensation threshold for the ES is $\sim$ 4 times higher than the threshold for the GS in Fig. 7(b). Both the GS and ES spectra show a series of week sidebands around the ZPL. The origin of these lines is discussed in the main text.

## IV. RF-induced Sidebands

### A. Raw sidebands data and correction

Figure 8(a) shows as measured high-resolution spectra of a trap GS as the function of RF power ($P_{RF}^2$) and fixed $F_{RF} \approx 7$ GHz. The imperfect transduction of RF power into acoustic leads to local heating of the sample. The latter is the origin of the redshift at higher $P_{RF}^2$. To simplify the analysis and presentation each spectrum was shifted horizontally by a small amount in order to match the energies of the zero-phonon line, cf. Fig. 8(b).

### B. Bessel-function fitting

Figure 9(a) displays a PL spectral map of the first excited state of a trap recorded for different amplitudes of the bulk acoustic wave (BAW) $A_{\rm BAW} = \sqrt{P_{\rm rf}}$. Here, $P_{\rm rf}$ is the nominal rf power applied to the BAWR. Individual PL spectra for different acoustic amplitudes are displayed by the symbols in the central panel. With increasing BAW amplitudes, one observes the appearance of an increasing number of well-defined sidebands displaced by multiples of the BAW quantum, $hf_{\rm BAW}$.

In the presence of a coherent harmonic driving of the BEC, the PL spectrum is expected to be proportional to $P[\omega]$:[3–5]

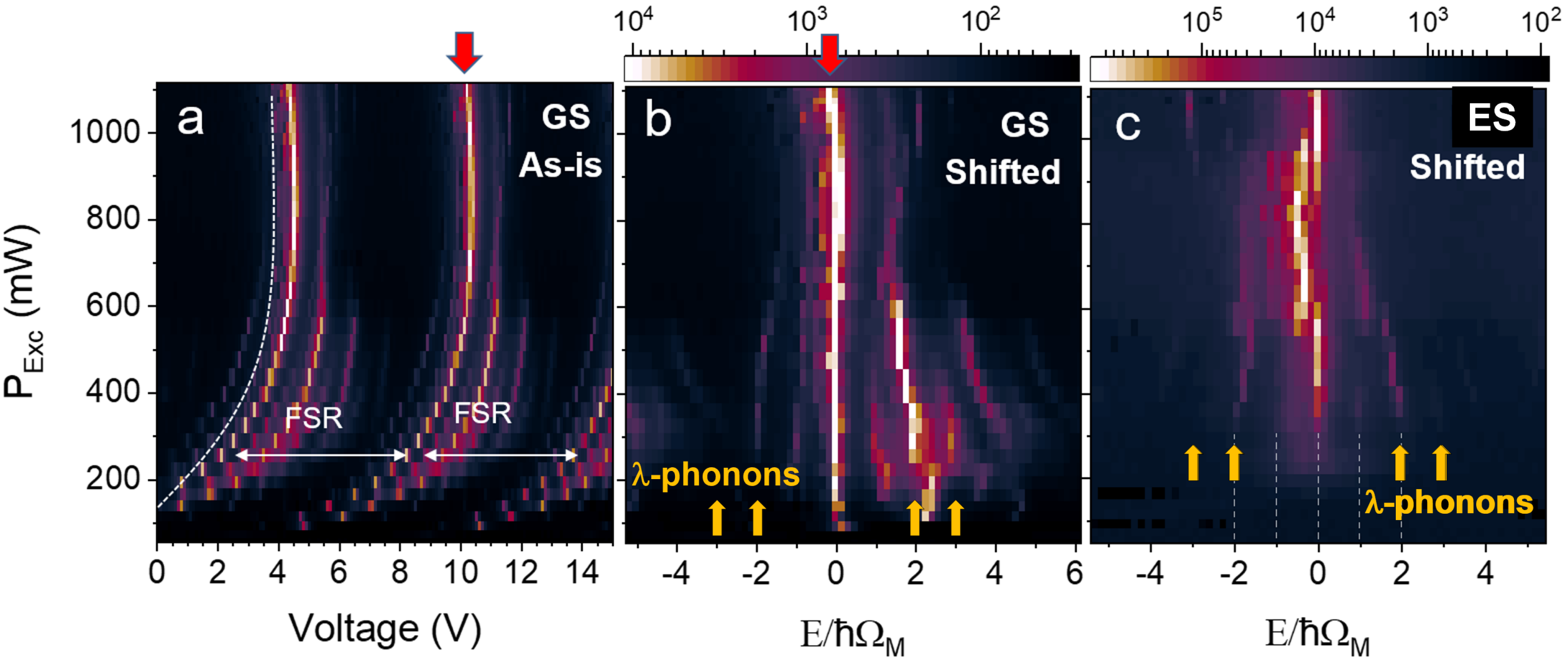

Fig. SM 7: Self-induced sidebands. (a) Raw maps of the ground state (GS) PL recorded for increasing optical excitation power $P_{\rm exc}$ focused on a spot of 10 $\mu$m positioned over a $4\times4$ $\mu m^2$ trap. The energy (horizontal axis) is specified in terms of the voltage applied to the piezo controlling the etalon with a voltage to frequency conversion factor of 6V per free spectral range of 68 GHz. The zero-phonon line (ZPL) is marked by a red arrow. (b) Same data after aligning all of the ZPL. The scale is stated in units of the LA $3\lambda$ phonon frequency $\hbar\Omega_{\rm M}$. (c) Shifted data for the excited state of the trap. The yellow arrows display the energies of the TA and LA $\lambda$-phonons.

$$P[\omega] = \sum_{n=-\infty}^{\infty} \frac{\Gamma}{2\pi} \frac{J_n^2(\chi)}{[\omega-(\omega_0-n\,\omega_a)]^2+(\frac{\Gamma}{2})^2}, \tag{1}$$

that is, a sum of Lorentzians with linewidths $\Gamma$, weighted by squared Bessel functions $J_n^2(\chi)$. $\chi$ is the amplitude modulation of the harmonic drive $\omega_0$ stated in units of the driving frequency $\omega_a$ (i.e., $\chi=\Delta\omega_0/\omega_a$). The Lorentzians have maxima at frequencies $\omega=\omega_0-n\omega_a$, where $n$ is an integer. Note that since $\sum_{n=-\infty}^{\infty} J_n^2(\chi)=1$, $\int_{-\infty}^{\infty} P[\omega]dw=1$ (see [functions.wolfram.com/Bessel-TypeFunctions/BesselJ/23/01/]), the effect of the modulation in Eq. 1 is thus to distribute the oscillator strength among the sidebands without affecting the overall PL intensity.

The solid lines in Figure 9(c,d,e) are fits of Eq. 1 to the experimental data. The latter yield the modulation amplitudes $\chi$ and linewidth $\Gamma$ displayed as a function of $A_{\rm BAW}$ in Fig. 9(f) and (g), respectively. $\chi$ is proportional to $A_{\rm BAW}$.

### C. RF-induced sidebands in the excited state

The experimental curves for the first excited state of the trap, shown in Figs. 9(c,d,e), are fitted well with a sum of Lorentzians with linewidths $\delta E$, weighted by squared Bessel functions $J_n^2(\chi)$, where $\chi$ – is the modulation amplitude, as described above.

Figure 9g shows the dependence of the sideband linewidth on the normalized acoustic amplitude ($A_{BAW}$). Similarly to the GS (see Fig. 2g in the main text), the linewidth $[\delta E(A_{BAW})]$ extracted from the fits decreases sharply by a factor of two from $\delta E(0.1)=0.22\hbar\Omega_M=1.75$ GHz to $\delta E(0.2)=0.1\hbar\Omega_M=0.8$ GHz and remains almost constant afterwards.

## V. Theoretical Background

This section theoretically analyses the coupling between confined phonons and polaritons, which can lead to self-oscillations in an intracavity trap. We start by describing the confined polariton and phonon in the trap and then proceed to the calculations of the interaction between them mediated by the deformation potential mechanism.

Both polaritons and phonons are tightly confined within the traps, which is assumed to have a thickness $m_z\lambda_{\rm BAW}/2$, where $\lambda_{\rm BAW}$ is the acoustic wavelength of the $3\lambda$ LA BAW. In order to describe the polariton and phonon confined fields, we use a reference frame with the $x$, $y$, and $z$ axes along the $[1\bar{1}0]$, $[110]$, and $[001]$ crystallographic directions, respectively. In the discussion to follow, a reference will also be made to the conventional cartesian frame with axes $x_c||[100]$, $y_c||[010]$, and $z_c||z||[001]$. Unless otherwise specified, the parameters used in the calculations are those listed in Table III.

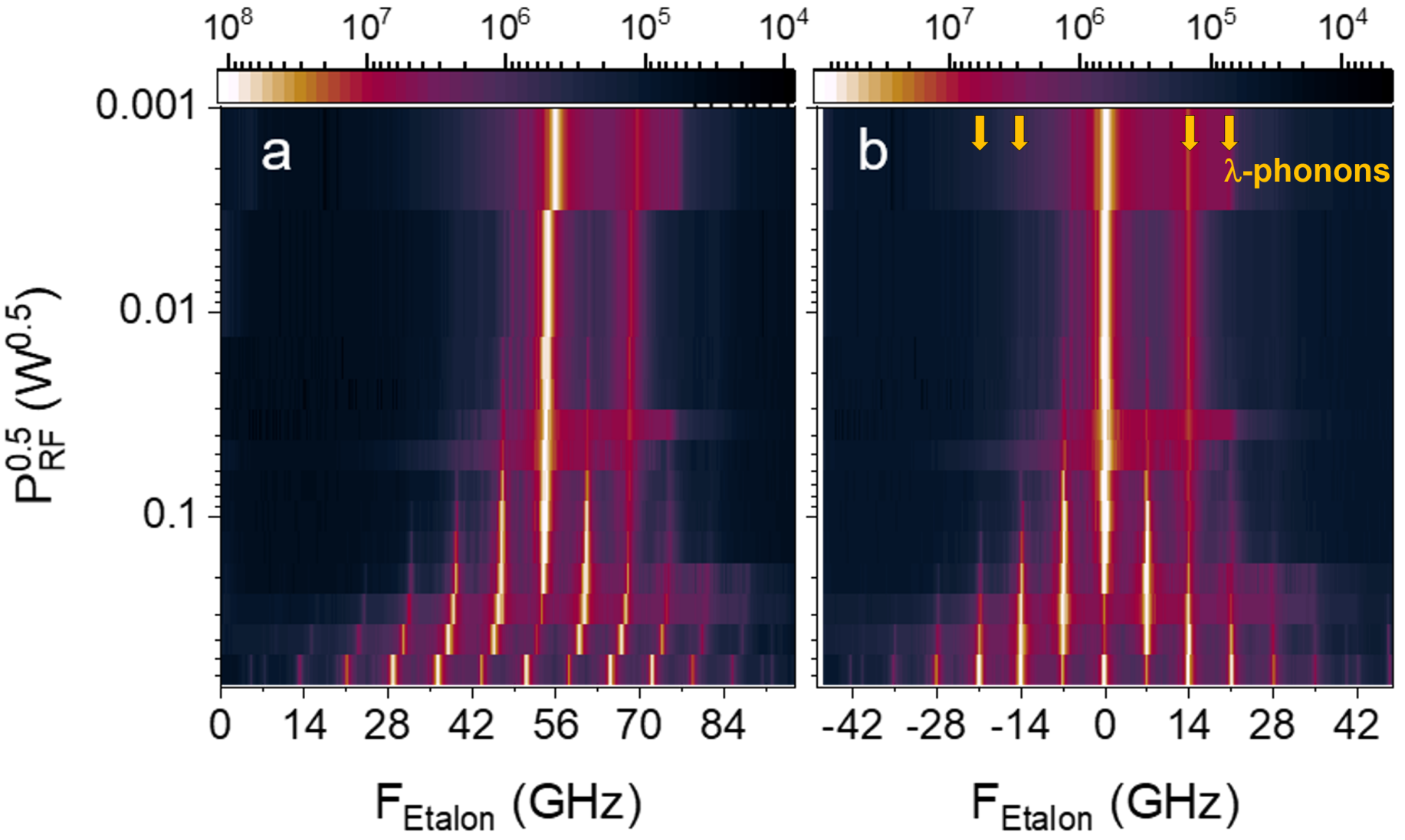

Fig. SM 8: Dependence on the RF power ($P_{RF}^2$) applied to the BAWR of the high-resolution PL spectrum of the polariton ground state (a) as-measured and (b) after energy correction to match the zero phonon line for the different $P_{RF}^2$. The yellow arrows display the energies of the TA and LA $\lambda$-phonons.

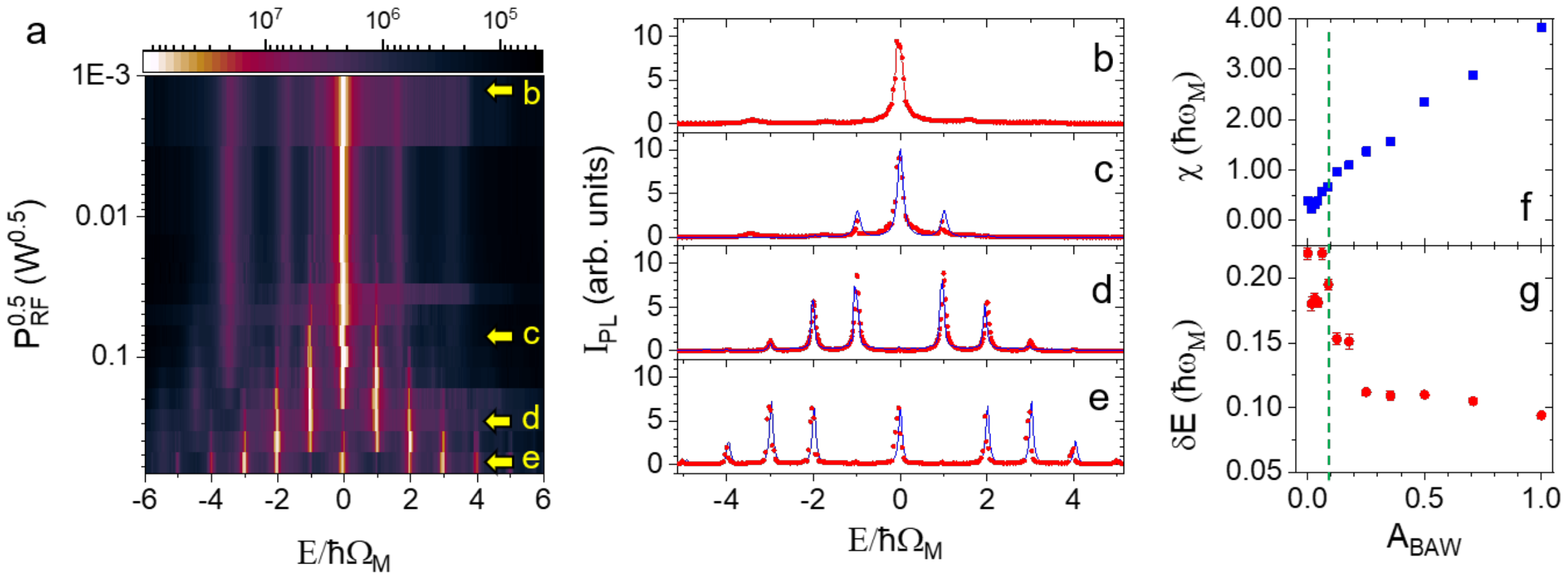

Fig. SM 9: (a) Dependence of the first excited state emission spectrum of a trap on acoustic amplitude $\sim \sqrt{P_{RF}}$. (b-e) PL spectra corresponding to the $\sqrt{P_{RF}}$ values indicated by yellow arrows (and letters) in panel a. The red dots are the data points while the blue solid lines are fits to Eq. 1. Dependence of the fitted modulation amplitude (f) and the linewidth (g) on the normalized acoustic amplitude. The vertical dashed green line in (f) and (g) indicates the acoustic amplitude for which the first sidebands appear.

In order to enhance the optomechanical coupling, the QWs (with a thickness of 15 nm much smaller than $\lambda_{\mathrm{BAW}}$) are placed close to an antinode of the light field, which is assumed to be at a position $z = 0$ [cf. Fig. 1(c)]. The MC was designed to ensure that the uniaxial strain along $z||[001]$ reaches its maximum amplitude $u_{zz,0}$ close to the same position [cf. Fig. 1(f)]. We will consider here only intra-cavity traps with a nominally square shape with the side length ($a$) in the $x$-$y$ plane. The MBE growth dynamics, however, distorts the trap geometry [2]. As a consequence, the lateral sizes of the trap will be taken to be $a_x = a + \Delta a/2$ and $a_y = a - \Delta a/2$ along the $x$ and $y$ directions, respectively, as illustrated in Fig. 10. The small size difference $\Delta a \ll a$ takes into account the anisotropic nature of the MBE overgrowth process on a structured surface, which yields traps with different sizes even for the overgrowth on a perfectly square mesa [2]. The trap dimension along the $z$ direction will be taken equal to the thickness of the MC spacer $d_s = m_z \frac{\lambda}{2n_s}$, where $m_z$ is an odd integer and $n_s$ is the average refractive index of the spacer layer.

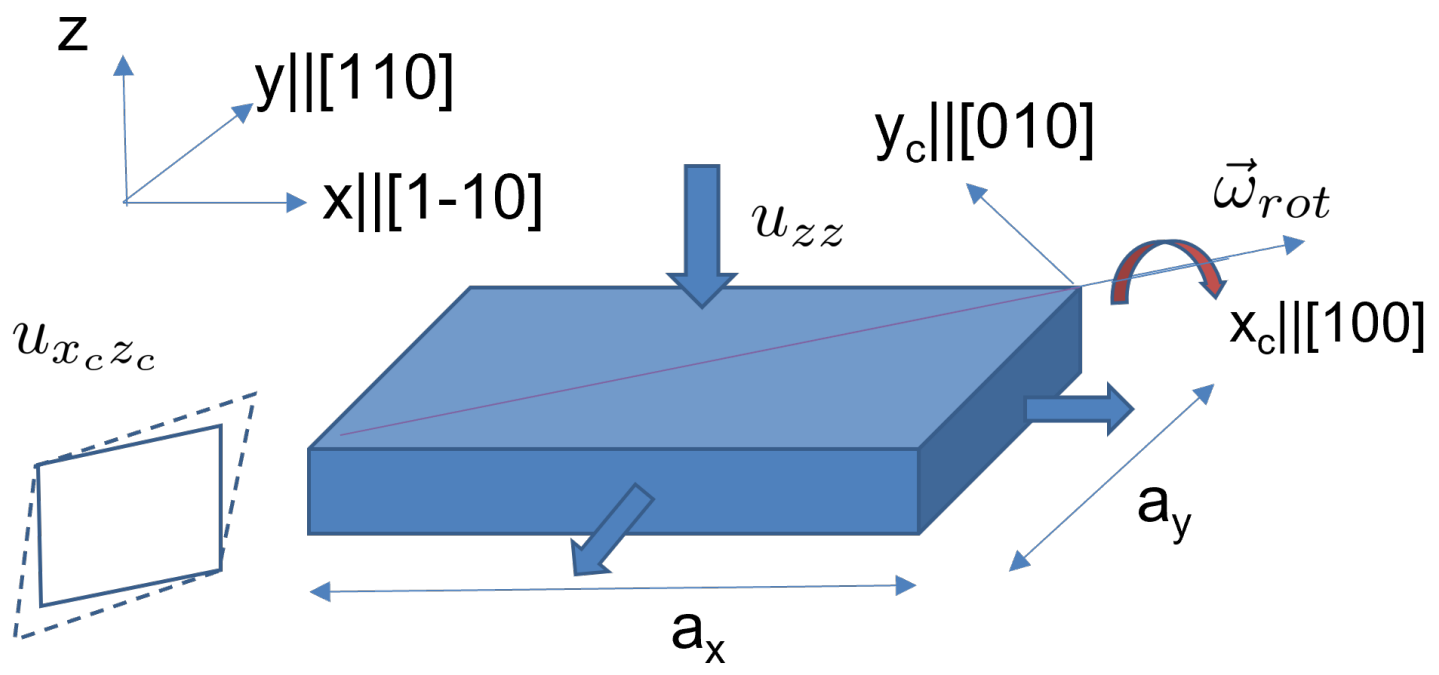

Fig. SM 10: Hybrid trap confining phonons and polaritons. The trap is assumed to have a rectangular shape with lateral dimensions $a_x = a + \Delta a/2$ and $a_y = a - \Delta a/2$, respectively, with $\Delta a \ll a$.

## A. Confined polariton modes

The envelope function of the confined polariton field in a trap can be written as:

$$\Psi^{(s)}_{(\mathrm{p_x p_y})}(x,y) = \sqrt{\frac{8k_x k_y}{3\pi^2}} \cos\left(k_x p_x x + \frac{1}{2}\pi(p_x - 1)\right) \cos\left(k_y p_y y + \frac{1}{2}\pi(p_y - 1)\right). \tag{2}$$

Here, $k_z = \pi/\lambda_{\mathrm{BAW}}$, $k_i = \pi/\ell_i$ $(i = x, y)$, and the indices $(p_i, p_j)$ (with $i = x, y$) are the number of the transverse lobes (i.e., perperdincular to z) of the polariton mode along the $x$ and $y$ directions, respectively. The previous equation applies for states with confinement energy shifts much smaller than the height of the trap potential barrier. The prefactor within the square root is a normalization factor ensuring that $|\langle\Psi^{(s)}_{(\mathrm{p_x p_y})}\rangle|^2 = 1$. As will be discussed in detail later, there are two polariton modes for each $(p_x p_y)$ pair labeled by the pseudo-spin superscript index $s$.

For small asymmetries $\Delta a/a$, the energy of the $(p_x p_y)$ polariton polariton state can be written as:

$$E^{(p_x p_y)}_{pol} = \frac{1}{3}(p_x^2 + p_y^2)E_{01} \tag{3}$$

where

$$E_{01} = \frac{3\hbar k_\perp^2}{2m_p}. \tag{4}$$

denotes the energy difference between the ground (GS) and the first excited state (ES) of the square potential. In the previous expression, $k_\perp = 2\pi/a$, and $m_p$ the polariton mass.

### *1. Confinement effects on excitonic states*

The high-resolution spectroscopic measurements on Fig. 6 shows a splitting of the polariton GS into two pseudo-spin levels (index $s$ in Eq. 2). This splitting calls for a more detailed description of the coupling between excitonic and photonic levels confined in the lateral trap potential. In the case of bare excitons, while the electron states have a simple s-type electronic wavefunction, the valence band states are superpositions of $p$-like $\langle X\rangle$, $\langle Y\rangle$, and $\langle Z\rangle$ orbitals, which can be mixed by the lateral confinement potential. The $\langle X\rangle$ and $\langle Y\rangle$ orbitals are degenerate for a square trap. This degeneracy can be lifted by a difference $\Delta a$ between the trap dimensions along the $x$ and $y$ directions according to:

$$\Delta_{\mathrm{XY}} = -2\left(\frac{\hbar^2 k_\perp^2}{2m_X}\right)\frac{\Delta a}{a} \approx -2\frac{m_{pol}}{m_X}E_{01}\frac{\Delta a}{a}, \tag{5}$$

where $m_X$ is the effective excitonic mass. For the right-most term of this equation, the factor $\left(\frac{\hbar^2 k_\perp^2}{2m_X}\right)$ has been expressed in terms of the energy difference $\Delta E_{01}$ between the first excited and ground polariton states of the trap.

Since $m_{pol} \ll m_X$, the lateral confinement shift of the exciton states can be neglected and $\Delta_{\rm XY}$ assumed to vanish. Furthermore, we will neglect fine splitting effects related to the exciton spin levels based on the fact that the light-matter interaction considerably increases the energetic splitting between these dark and bright exciton levels.

### 2. *Confinement effects on the photonic states*

We now introduce the bare photonic states and their coupling to the excitons. The bare states will be described in a basis $\left(L_R = \frac{1}{\sqrt{2}}(L_{\rm TM} + iL_{\rm TE}), L_L = \frac{1}{\sqrt{2}}(L_{\rm TM} - iL_{\rm TE})\right)$ of right ($L_R$) and left circularly polarized light ($L_L$) propagating with a wave vector component $k_\perp \propto \sin\theta_L$ perpendicular to the $z$-axis. $\theta_L$ is the propagation angle between the light beam and the $z$ axis, while $L_{\rm TM}$ and $L_{\rm TE}$ are the corresponding light modes with polarization in the $x-z$ plane (TM) and perpendicular to it (TE), respectively. Due to the dependence of the light propagation through the DBRs on polarization direction, the $L_{\rm TM}$ and $L_{\rm TE}$ eigenstates are, in general, non-degerate for $\theta_L \neq 0$ [6]. The energy of the bare photon states in a trap can be described by the following Hamiltonian in the basis $(L_{\rm TM}, L_{\rm TE})$ [7]:

$$
\begin{aligned}
H_{LT}^{(p_x p_y)}(k_{av}) &= \frac{\hbar^2 k_{av}^2}{2m_{ph}} + \frac{\Delta_{\rm LT}(k_{av})}{k_{av}^2}\left[\begin{array}{c}(p_x k_x)^2 - (p_y k_y)^2 \\ 2p_x k_x p_y k_y\end{array}\right]_{av} \\
&= \frac{2}{3}\left(p_x^2 + p_y^2\right) E_{01} + \delta_{\rm LT}^{(p_x p_y)} \qquad (6) \\
k_{av}^2 &= (p_x k_x)^2 + (p_y k_y)^2 \\
\delta_{\rm LT}^{(p_x p_y)} &= (\hbar\omega_{\rm TM} - \hbar\omega_{\rm TE}) \approx \Delta_{\rm LT}(k_{av}, 0)\left[\frac{p_x^2 - p_y^2}{p_x^2 + p_y^2} - 2\frac{p_x p_y}{p_x^2 + p_y^2}\frac{\Delta a}{a}\right]. \qquad (7)
\end{aligned}
$$

where $m_{ph}$ is the bare photon mass. In the second line of Eq. 6, the confinement energy shift of the bare photonic level is expressed in terms of the polariton interlevel splitting (in a manner analogous to the excitonic states in Eq. 5) assuming that the bare photonic mass in the MC is half the polariton mass $m_p$ (i.e., $m_{ph} \approx m_p/2$).

The longitudinal-transverse splitting $\Delta_{\rm LT}(k_{av}, 0)$ depends on the photon wave vector as well as on the layer structure of the MC. When averaged over the wave function of a polariton mode $(p_x p_y)$, the energy splitting between the $L_{\rm TM}$ (with energy $\hbar\omega_{\rm TM}$) and $L_{\rm TE}$ ($\hbar\omega_{\rm TE}$) states becomes equal to $\delta_{\rm LT}^{(p_x p_y)}$ given by Eq. 7. This approximation on the right-hand side applies for $\delta a/a \ll 1$.

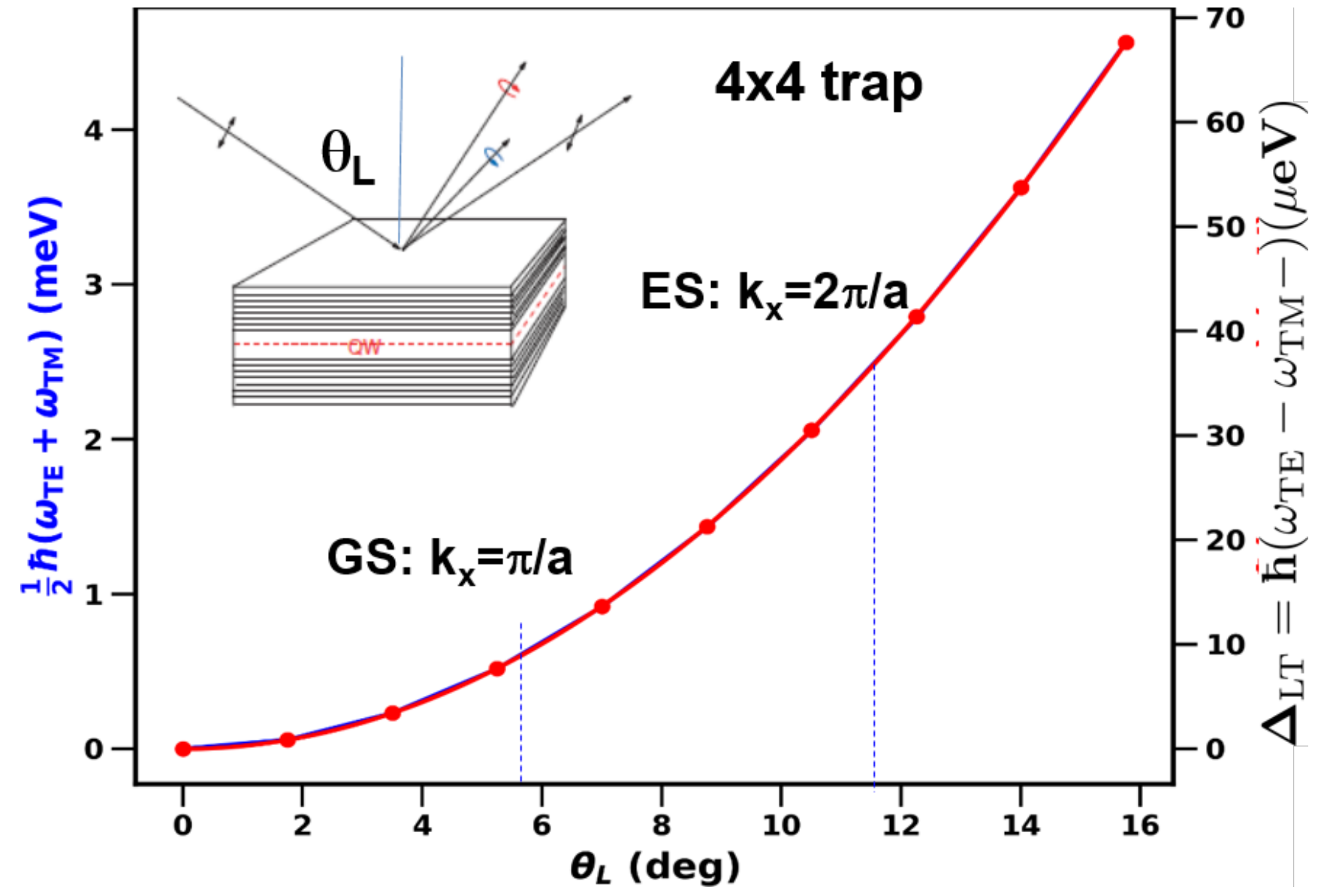

Fig. SM 11: Calculated average energy $((\hbar\omega_{\rm TE} + \hbar\omega_{\rm TM})/2$, left vertical scale) as well as the splitting $\delta_{\rm TE-TM} = \hbar(\omega_{\rm TE} - \omega_{\rm TM})$ between transverse electric (TE) and transverse magnetic (TM) modes of the optical microcavity of Table I as a function of the angle of incidence $\theta_L$ (relative to the $z$ axis) in the $x-z$ plane. The dashed vertical line marks $\theta_L$ corresponding to an in-plane wave vector $\vec{k} = (k_{\rm x}, k_{\rm y}) = (\pi/a, 0)$ for the $(p_x, p_y) = (1, 1)$ polariton mode yielding a splitting of 9.5 $\mu$eV, where $a = 4$ $\mu$m is the trap size.

The pseudo-spin splitting $\Delta_{\rm LT}$ in Eq. 7 corresponds to the splitting between the bare photonic states for light propagation with an in-plane wave vector $k_x$. Figure 11 displays the average energy $(\hbar\omega_{\rm TM}(\vec{k})+\hbar\omega_{\rm TE}(\vec{k}))/2$ as well as $\Delta_{\rm LT}(\vec{k})=(\hbar\omega_{\rm TM}(\vec{k})-\hbar\omega_{\rm TE}(\vec{k}))$ calculated for the $(p_xp_y)=(1,1)$ polariton mode with a wave vector $\vec{k}=(k_x,0)$ in the sample structure used in this work. The calculations were carried out using a transfer matrix approach to determine the energy of the light states in the $x--z$ plane with polarization along $y$ (TE-polarization) and in the incidence plane (TM-polarization) in an empty MC, i.e., without including excitonic contributions to the dielectric response. For small propagation angles, both the average energy and the pseudo-spin splitting increase quadratically with the wave vector. The dashed vertical line marks the wave vector component $k_x=\pi/a_x$ corresponding to the polariton GS of a $4\times 4$ $\mu$m$^2$ trap. From the splitting $\Delta_{\rm LT}(k)=9.5$ $\mu$eV indicated by the dashed line, one obtains from Eq. 7 energy splittings $\delta^{(1,1)}_{\rm LT}=0.48$ $\mu$eV and $\delta^{(2,1)}_{\rm LT}=3.6$ $\mu$eV for the GS and first excited state (ES) of a trap with $\Delta a/a=10\%$. In both cases, the splitting is much smaller than the $3\lambda$ LA phonon quantum of $\hbar\omega_m=28$ $\mu$eV.

## B. Confined phonon modes

Similarly to the case of polaritons, we will assume the displacement field $\vec{u}(x,y,z)$ of the phonons to be confined within a box with lateral sizes corresponding to the trap size and dimension along $z$ equal to the thickness of the cavity spacer $m_s\lambda_{\rm BAW}/2$, as illustrated in Fig. 10.

The confined phonon modes at the QW plane can be approximated by those of a rectangular box with dimensions $a_x\sim a_y$ along the $x$ and $y$ directions, respectively. If the lateral surfaces of the box at $x=\pm a_x/2$ and $y=\pm a_y/2$ are free to move, the displacement field $\vec{u}(x,y,z)$ for the three eigenmodes is $\left(\vec{u}^{TA'}_{(\rm m_xm_y)},\vec{u}^{TA}_{(\rm m_xm_y)},\vec{u}^{LA'}_{(\rm m_xm_y)}\right)$ with a wave vector $k_z=m_z\pi/\lambda_{\rm BAW}$ along $z||[001]$. The indices $(m_xm_y)$, $m_i=1,2,\ldots,i=x,y$ describe the number of the transverse lobes (i.e., perperdincular to $z$) of the acoustic mode. $m_z$ is the index of the longitudinal mode being equal to 1 and 3 for the $\lambda$- and $3\lambda$ phonon modes, respectively. The displament field of these modes can be stated as the columns of the following matrix:

$$\begin{aligned}\vec{u}_{(\rm m_xm_y)}(x,y,z) &= \left(\vec{u}^{TA'}_{(\rm m_xm_y)},\vec{u}^{TA}_{(\rm m_xm_y)},\vec{u}^{LA'}_{(\rm m_xm_y)}\right) \\ &= u_{zz_0}\begin{pmatrix}\frac{r_T}{k_xm_x} & \frac{1}{k_xm_x} & -\frac{k_xm_x}{2k_z^2r_T}\\ \frac{r_T}{k_ym_y} & -\frac{1}{k_ym_y} & -\frac{k_ym_y}{2k_z^2r_T}\\ \frac{1}{k_z} & 0 & \frac{1}{k_z}\end{pmatrix} e^{i(k_xm_xx+k_ym_yy+k_zz)},\end{aligned} \tag{8}$$

where

$$r_T=c_{12}/(c_{11}+c_{12}). \tag{9}$$

is the ratio between the uniaxial deformation perpendicular and parallel to the $z$ axis. For GaAs, $r_T=-0.31$. The amplitude of the acoustic field in Eq. 8 is stated in terms of the amplitude $u_{zz,0}$ of the uniaxial strain along $z$, which, as previously mentioned, has a maximum close to the QW position.

The $\vec{u}^{TA}_{(\rm m_xm_y)}(x,y,z)$ eigenmode in Eq. 8 is a pure transversal acoustic (TA) mode polarized in the $x$-$y$ plane of the trap and angular frequency $f_{TA}\sim\frac{1}{2\pi}\sqrt{\frac{c_{44}}{\rho}}k_z$. The two other modes have mixed transversal and longitudinal acoustic (LA) character induced by the lateral confinement. The mixing is dictated by $r_T$ as well as by the dimensions of the trap. For large traps (i.e., $a\gg\lambda_{\rm BAW}$), $\vec{u}^{TA'}_{(\rm m_xm_y)}(x,y,z)$ can be well approximated by a TA wave with a small displacement component along $z$. Its angular frequency is slighly smaller than $f_{TA}$. $\vec{u}^{LA'}_{(\rm m_xm_y)}(x,y,z)$ is essentially a LA mode with a small transverse component and angular frequency $f_{LA}\sim\frac{1}{2\pi}\sqrt{\frac{c_{11}}{\rho}}k_z$.

The frequency of the $3\lambda$ phonon modes for different $(m_xm_y)$ combinations are listed in the first two columns of Table V. These frequencies were determined using the elastic properties of GaAs. Since the spacer of the MC also includes (Al,Ga)As layers, which have higher acoustic velocities, the listed values sligthly underestimate the measured resonance frequencies. The corresponding frequencies for the $\lambda$ phonons are three times as large. To a very good approximation (i.e., with deviations less than $\sim 5\%$), the relationship between the mode frequencies is $f_{TA'}\approx f_{TA}\approx 0.7f_{\rm LA}$ for modes with $m_i\leq 2$ $(i=1,2)$. It is interesting to note that the frequencies $f_{TA'}$ and $f_{TA}$ for the $\lambda$ phonons are approximately twice as large as the frequency $f_{LA}$ for the $3\lambda$ phonons. By solving the elastic equations for a $4\times 4$ $\mu$tm$^2$ trap, we obtain the following frequency ratios:

$$f^{(\lambda)}_{TA'}\approx f^{(\lambda)}_{TA}=2.07f^{(3\lambda)}_{LA}=2.07f_{\rm BAW} \tag{10}$$

$$f^{(\lambda)}_{LA'}=2.95f^{(\lambda)}_{LA}=2.95f_{\rm BAW}, \tag{11}$$

TABLE III: Parameters used in the calculations.

| Parameter | Value | Unit | Remarks |
|---|---|---|---|
| Material properties | | | unless specified, for GaAs |
| $a_h$ | -9 | eV | hydrostatic deformation potential [8] |
| $b$ | -2 | eV | uniaxial deformation potential [8] |
| $d$ | -6 | eV | shear deformation potential [8] |
| $c_{11}^{(c)}$ | $1.183 \times 10^{11}$ | $N/m^2$ | elastic constant [9] |
| $c_{12}^{(c)}$ | $0.532 \times 10^{11}$ | $N/m^2$ | elastic constant [9] |
| $c_{44}^{(c)}$ | $0.595 \times 10^{11}$ | $N/m^2$ | elastic constant [9] |
| $\rho$ | 5316.5 | $kg/m^3$ | density [9] |
| $\epsilon_{rel}$ | 12 | - | static dielectric constant |
| $e_{14}$ | -0.16 | $C/m^2$ | piezoelectric strain coefficient [9] |
| Trap properties | | | |
| $a_{\text{trap}}$ | 4 | $\mu$m | trap size |
| $\Delta a_{\text{trap}}$ | - | $\mu$m | trap anisotropy (cf. Fig. 10) |
| Polariton properties | | | |
| $\delta_{\text{CX}}$ | 0 | meV | Cavity-exciton detuning |
| $E_{01}$ | 1 | meV | splitting between the polariton GS and first ES |
| $\Delta_{\text{XY}}$ | 100 | $\mu$eV | Trap anisotropic energy (real part) |
| $\Delta_R$ | 0.0 | eV | Trap anisotropy energy (imag. part) |
| $\Delta_{\text{LT}}$ | 9.5 | $\mu$eV | LT splitting for the ground polariton $(p_x p_y) = (1,1)$ state |
| $\delta_{\text{LT}}^{(1,1)}$ | 1.5 | $\mu$eV | integrated LT splitting for the $(p_x p_y) = (1,1)$ state, cf. Eq. (7) |
| $\Delta_{\text{hh}-\text{lh}}$ | 6 | meV | QW lh-hh energy splitting |
| $\Omega_{\text{R}}$ | 3 | meV | polariton Rabi coupling for e-hh states |
| $\Gamma_{\text{pol}}$ | 4.8 | $\mu$eV | Polariton BEC decoherence rate |
| $N_{\text{pol}}$ | $10^5$ | | Polariton BEC population at the threshold |
| Phonon properties | | | |
| $f_{\text{mec}}$ | 7 | GHz | Frequency of piezoelectrically generated phonons |
| $\Gamma_{\text{mec}}$ | $5.5 \times 10^{-3}$ | $\mu$eV | Phonon decoherence rate |

where $f_{\text{BAW}}$ is the frequency of the phonons generated by the BAWRs. The $\lambda$-phonon frequencies are indicated by thick arrows in Figs. 7, 8 and 9 of this supplement.

The vibrational eigenmodes of the trap are superpositions of the modes of Eq. 8 satisfying the boundary conditions at the trap borders. Here, we will assume that the uniaxial strain components vanish at the lateral trap borders. In this case, we obtain the following expression for the displacement field of mode $\vec{u}^{LA'}_{(\text{m}_\text{x}\text{m}_\text{y})}(x,y,z)$, which is also the one that can be efficiently excited by the BAWRs (similar expression can be easily derived for the other modes from Eq. 8):

$$\vec{u}_{(\text{m}_\text{x}\text{m}_\text{y})}(x,y,z) = \vec{u}^{LA'}_{(\text{m}_\text{x}\text{m}_\text{y})}(x,y,z) = \begin{pmatrix} \frac{k_x m_x \cos(k_z z)\cos\left(\frac{1}{2}m_x(\pi-2k_x x)\right)\sin\left(\frac{1}{2}m_y(\pi-2k_y y)\right)}{2k_z^2 r_T} \\ \frac{k_y m_y \cos(k_z z)\sin\left(\frac{1}{2}m_x(\pi-2k_x x)\right)\cos\left(\frac{1}{2}m_y(\pi-2k_y y)\right)}{2k_z^2 r_T} \\ -\frac{\sin(k_z z)\sin\left(\frac{1}{2}m_x(\pi-2k_x x)\right)\sin\left(\frac{1}{2}m_y(\pi-2k_y y)\right)}{k_z} \end{pmatrix} u_{zz_0} \tag{12}$$

The changes in the lattice induced by the phonon field are proportional to the displacement gradient $\nabla\vec{u}_{(m_x m_y)}(x,y)$. The latter can be decomposed into a symmetric ($\varepsilon_{m_x m_y} = \frac{1}{2}\left[\nabla u_{(m_x m_y)} + (\nabla u_{(m_x m_y)})^T\right]$) and an anti-symmetric ($\vec{\omega}_{(m_x m_y)} = \frac{1}{2}\left[\nabla u_{(m_x m_y)} - (\nabla u_{(m_x m_y)})^T\right]$) contribution. The first corresponds to the strain field while the second yields a pure rotation of the lattice (i.e., without the lattice distortion). $\vec{\omega}_{(m_x m_y)}$ is much smaller than $\varepsilon_{m_x m_y}$ and will be neglected in the next sections.

The amplitude of the strain field $u_{zz,0}$ in Eq. 12 can be determined by calculating the energy stored in the strain field. This amplitude $u_{zz,0}$ can be stated in terms of the phonon frequency $\hbar\omega_{\text{mec}}$, and number $n_{\text{mec}}$, and the zero-point motion $u_{zz,0}^{ZPM}$ as:

$$u_{zz,0} = \sqrt{n_{\rm mec}} u_{zz,0}^{ZPM}, \quad u_{zz,0}^{ZPM} = \frac{64\sqrt{f_{\rm BAW}}}{a_x a_y} \sqrt{\frac{1}{128\rho v_{LA}^3 + \frac{8\rho v_{LA}{}^7}{(a_x a_y)^4} f_{\rm BAW}^4 r_T{}^2}}. \tag{13}$$

The second term in the denominator within the square root is the correction introduced by the reduced trap size.

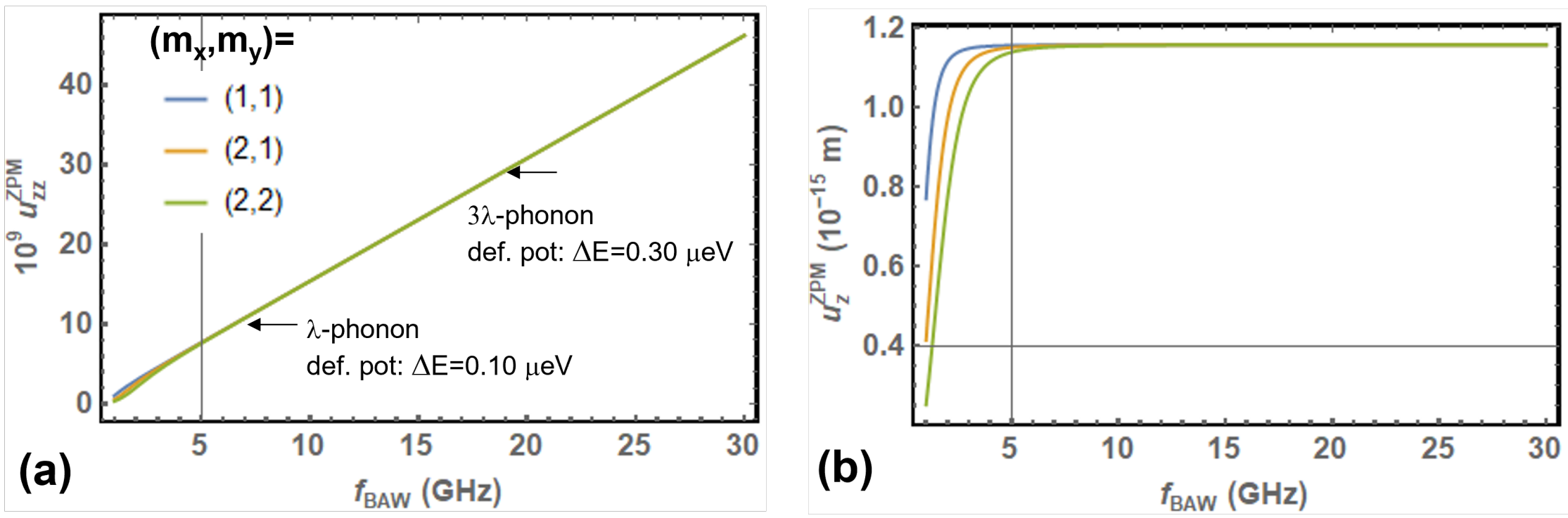

Fig. SM 12: Calculated amplitude of the (a) strain field component (a) $u_{zz}^{\rm ZPM}$ and of (b) the $z$-oriented displacement $u_z^{\rm ZPM}$ corresponding to the zero-point fluctuation of a phonon with indices $m_x m_y$ in a $4 \times 4$ $\mu$m$^2$ trap with different thicknesses $\lambda_{\rm BAW}/2 = xx$, which yield the phonon frequencies displayed by the horizontal axis.

Figures 12(a) and 12(a) display the amplitude of the strain ($u_{zz}^{\rm ZPM}$) and displacement field component ($u_z^{\rm ZPM}$), respectively, calculated using Eq. 13 for a single phonon confined in a $4 \times 4$ $\mu$m$^2$ trap with different thicknesses $\lambda_{\rm BAW}/2$ (and, thus different frequencies $f_{\rm BAW}$). For short phonon wavelengths (i.e., frequencies above $\sim 5$ GHz) all modes indexed by $(m_x m_y)$ have approximately the same frequency dependence with the strain (displacement) field proportional to $f_{\rm BAW}^{1/2}$ ($1/f_{\rm BAW}^{1/2}$). For a 7 GHz phonons investigated in this work, the strain amplitudes $u_{zz}^{\rm ZPM} \sim 10^{-8}$ are expected to induce energy shifts of the excitonic electron-heavy hole transition of approximately $a_h u_{zz}^{\rm ZPM} \approx 0.10$ $\mu$eV$= 2\pi\hbar$ 25 MHz, where $a_h$ is the hydrostatic exciton deformation potential (cf. Table III).

## C. Phonon-polariton coupling

### 1. *Coupling mechanisms*

The phonon field acts on the polariton states in the three different ways illustrated schematically in Fig. 13. The solid and dashed lines for a polariton with orbital index $p_x p_y$ indicate the states with up (↑) and down (↓) pseudo-spin indices, respectively:

- The phonons can modulate the energy of the individual polariton states with the coupling strength $\delta E_{hh,(m_x m_y)}^{(p_x p_y)}$: if the energy modulation frequency exceeds the polariton linewidth, this modulation leads to the formation of emission sidebands.
- The phonons can mix the polariton pseudo-spin states $\Psi_{(p_x p_y)}^{(s)}(x, y)$ with the same orbital indices. This *intra-mode coupling* process is quantified by the coupling energy $g_{\uparrow\downarrow(m_x m_y)}^{(p_x p_y)}$).
- Alternatively, they can induce an *inter-mode coupling* between polaritons with different orbital indices. The coupling strengths for inter-mode spin-conserving and spin-flipping processes are denoted by $g_{\uparrow\uparrow(m_x m_y)}^{(p_x p_y)\leftrightarrow(p'_x p'_y)} = g_{\downarrow\downarrow(m_x m_y)}^{(p_x p_y)\leftrightarrow(p'_x p'_y)}$ and $g_{\uparrow\downarrow(m_x m_y)}^{(p_x p_y)\leftrightarrow(p'_x p'_y)}$ factors, respectively.

Details of the coupling mechanisms will be addressed in the next sections.

### 2. *Averaged strain field for polariton coupling*

We will restrict ourselves to phonon-polariton coupling relying on the deformation potential mechanism, which has been shown to dominate the coupling in (Al,Ga)As MCs [1]. In this framework, the effects of the acoustic field on the coupling of polariton mode $\Psi_{(p_x p_y)}^{(s)*}$ are mediated by its excitonic component with (orthonormalized) wave function $\chi^{(s)*}_{(p_x p_y)}$, which we will assume to have the same spatial dependence as the polariton wave function $\Psi^{(s)*}_{(p_x p_y)}$ in Eq. 2. The coupling elements can be stated as:

$$\langle \chi^{(s')*}_{(p'_x, p'_y)} | \nabla \vec{u}_{(m_x m_y)}(x,y) | \chi^{(s)}_{(p_x p_y)} \rangle = \varepsilon^{(p'_x, p'_y)\leftrightarrow(p_x p_y)}_{(m_x m_y)} + \vec{\omega}^{(p'_x, p'_y)\leftrightarrow(p_x p_y)}_{(m_x m_y)} \tag{14}$$

$$\varepsilon^{(p'_x, p'_y, s')\leftrightarrow(p_x p_y, s)}_{(m_x m_y)} = \int \varepsilon_{(m_x m_y)} \chi^{(s')*}_{(p'_x, p'_y)}(x,y) \chi^{(s)}_{(p_x p_y)}(x,y) dxdy \tag{15}$$

$$\vec{\omega}^{(p'_x, p'_y, s')\leftrightarrow(p_x p_y, s)}_{(m_x m_y)} = \int \vec{\omega}_{(m_x m_y)} \chi^{(s')*}_{(p'_x, p'_y)}(x,y) \chi^{(s)}_{(p_x p_y)}(x,y) dxdy. \tag{16}$$

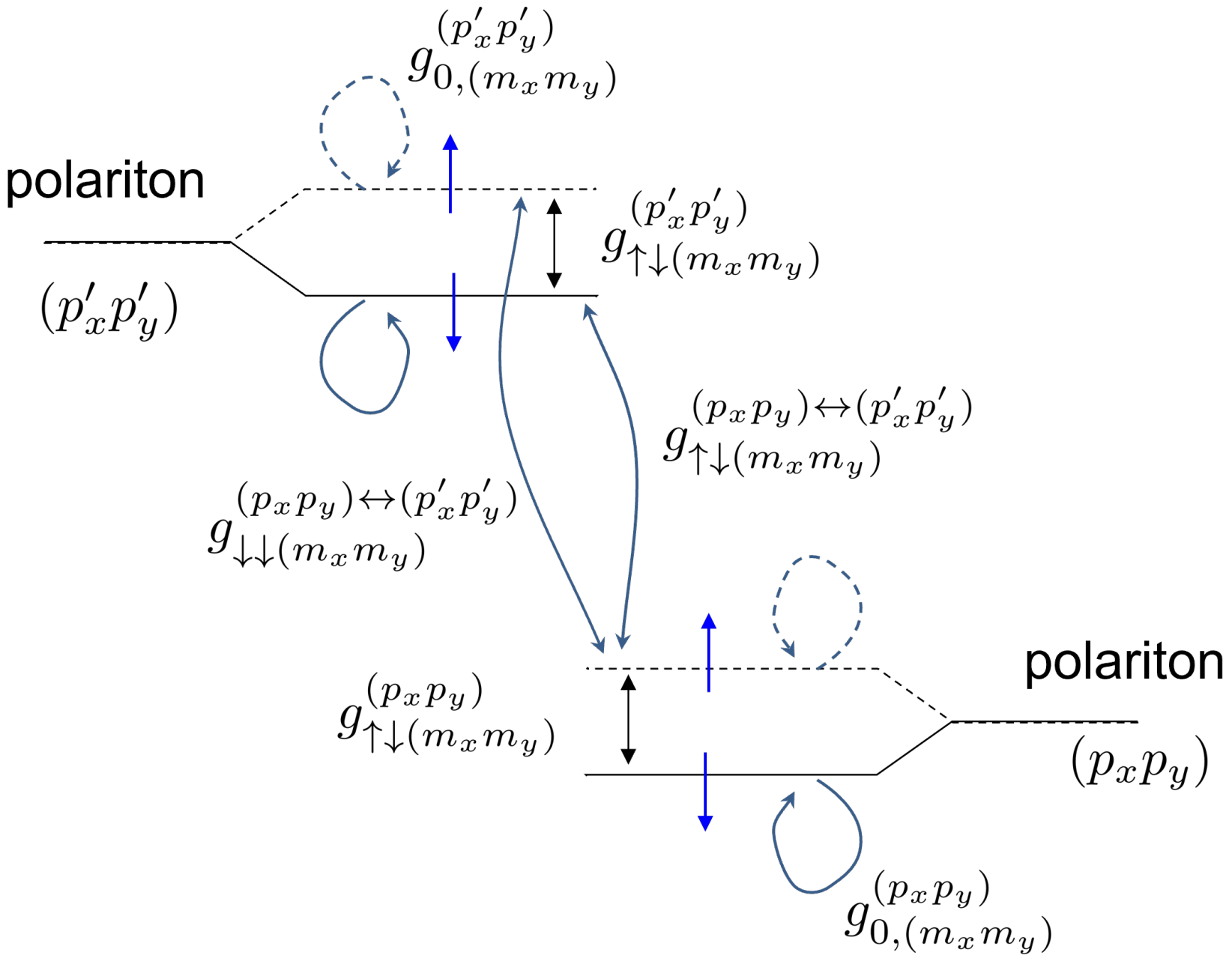

Fig. SM 13: Coupling between the polariton levels $(p_x p_y)$ and $(p'_x, p'_y)$ mediated by a $(m_x m_y)$ phonon. Intra-level coupling induces the energy modulation of the pseudo-spin levels (quantified by the coupling strength $g^{(p_x p_y)}_{0,(m_x m_y)}$ and $g^{(p'_x p'_y)}_{0,(m_x m_y)}$) as well as coupling between the pseudo-spin states ($g^{(p_x p_y)}_{\uparrow\downarrow(m_x m_y)}$ and $g^{(p'_x p'_y)}_{\uparrow\downarrow(m_x m_y)}$). The inter-level coupling is mediated by the spin-conserving and non-conserving factors $g^{(p_x p_y)\leftrightarrow(p'_x p'_y)}_{\uparrow\uparrow(m_x m_y)} = g^{(p_x p_y)\leftrightarrow(p'_x p'_y)}_{\downarrow\downarrow(m_x m_y)}$ and $g^{(p_x p_y)\leftrightarrow(p'_x p'_y)}_{\uparrow\downarrow(m_x m_y)}$, respectively.

From this point on, we will implicitly assume that the optoelectronic coupling is mediated by the excitonic component and no longer distinguish between $\chi^{(s)*}_{(p_x p_y)}$ and $\Psi^{(s)*}_{(p_x p_y)}$. It can be shown that the right-hand side of Eq. 15, which corresponds to the antisymmetric part related to rotation, vanishes when averaged over the excitonic wavefunction of the quasi-LA mode of Eq. 12.

We will be first interested in the effects of the strain from the $\vec{u}^{LA'}_{(\mathrm{m_x m_y})}$ confined phonons on the lower lying polariton levels corresponding to the pseudo-spin split ground state (GS) $(p_x p_y) = (1,1)$ and first excited states (ES) with $(p_x p_y) = (2,1)$, $(1,2)$, or $(2,2)$. These modes are expected to have the longest coherence time and smallest energy separations, which can be comparable to the phonon quantum. The relevant phonons are those with the lowest energy, which are well confined within the cavity [i.e., the $(m_x m_y) = (1,1)$ GS mode and the first ES modes with $(m_x m_y) = (2,1)$ or $(1,2)$].

Expressions can be derived from Eqs. 12 and 15 for the strain field components $\varepsilon^{(p'_x, p'_y, s')\leftrightarrow(p_x p_y, s)}_{(m_x m_y)}$ induced by a $(m_x m_y)$ phonon and averaged over the polariton modes $(p'_x, p'_y)$ and $(p_x p_y)$. As an example, Table IV summarizes the average strain fields associated with GS phonon mode $(m_x m_y) = (1,1)$. The strain tenson is listed in engineering notation with $\varepsilon^c_{(m_x m_y)} = \left(e^c_{xx}, e^c_{yy}, e^c_{zz}, e^c_{yz}, e^c_{xz}, e^c_{xy}\right)$ in the cartesian reference frame (superscript $^c$). We use the representation in the cartesian frame to obtain a more familiar representation of the Pikus and Bir Hamiltonian in Sec. V C 3.

TABLE IV: Integrated effective strain $\varepsilon^{(p'_x,p'_y,s')\leftrightarrow(p_xp_y,s)}_{(m_xm_y)} = \left(e^c_{xx}, e^c_{yy}, e^c_{zz}, e^c_{yz}, e^c_{xz}, e^c_{xy}\right)$ in units of $\left(\frac{u_{zz}}{\pi^2}\right)$ associated with the phonon mode $\vec{u}^{LA'}_{(\mathrm{m_x m_y})}$ with $(m_x = 1, m_y = 1)$. The numerical values were calculated for a $4 \times 4$ $\mu$m$^2$ trap with a shape anisotropy $\Delta a/a = 0.1$. Entries marked as "tr" in the matrix are equal to their transpose.

| $(p_xp_y)$ | $(p'_xp'_y) = (1,1)$ | $(p'_xp'_y) = (1,2)$ | $(p'_xp'_y) = (2,1)$ | $(p'_xp'_y) = (2,2)$ |
|---|---|---|---|---|
| 11 | $\begin{pmatrix} -\frac{128r_T}{27} & 0 & \frac{64k_\perp^2}{27k_z^2r_T} \\ -\frac{128r_T}{27} & 0 & \frac{64k_\perp^2}{27k_z^2r_T} \\ -\frac{128}{27} & 0 & -\frac{128}{27} \\ 0 & 0 & 0 \\ 0 & 0 & 0 \\ 0 & \frac{256}{27} & \frac{128k_\perp^2}{27k_z^2r_T}\frac{\Delta a}{a} \end{pmatrix}$ | $tr$ | $tr$ | $tr$ |
| 12 | 0 | $\begin{pmatrix} -\frac{512r_T}{135} & 0 & \frac{256k_\perp^2}{135k_z^2r_T} \\ -\frac{512r_T}{135} & 0 & \frac{256k_\perp^2}{135k_z^2r_T} \\ -\frac{512}{135} & 0 & -\frac{512}{135} \\ 0 & 0 & 0 \\ 0 & 0 & 0 \\ 0 & \frac{1024}{135} & \frac{512k_\perp^2}{135k_z^2r_T}\frac{\Delta a}{a} \end{pmatrix}$ | $tr$ | $tr$ |
| 21 | 0 | $\begin{pmatrix} \frac{\pi^2r_T}{6} & \frac{\pi^2}{6}\frac{\Delta a}{a} & -\frac{\pi^2k_\perp^2}{12k_z^2r_T} \\ -\frac{1}{6}\pi^2r_T & -\frac{\pi^2}{6}\frac{\Delta a}{a} & \frac{\pi^2k_\perp^2}{12k_z^2r_T} \\ 0 & 0 & 0 \\ 0 & 0 & 0 \\ 0 & 0 & 0 \\ 0 & 0 & 0 \end{pmatrix}$ | $\begin{pmatrix} -\frac{512r_T}{135} & 0 & \frac{256k_\perp^2}{135k_z^2r_T} \\ -\frac{512r_T}{135} & 0 & \frac{256k_\perp^2}{135k_z^2r_T} \\ -\frac{512}{135} & 0 & -\frac{512}{135} \\ 0 & 0 & 0 \\ 0 & 0 & 0 \\ 0 & \frac{1024}{135} & \frac{512k_\perp^2}{135k_z^2r_T}\frac{\Delta a}{a} \end{pmatrix}$ | 0 |
| 22 | $\begin{pmatrix} \frac{\pi^2r_T}{6} & \frac{\pi^2}{6}\frac{\Delta a}{a} & -\frac{\pi^2k_\perp^2}{12k_z^2r_T} \\ -\frac{1}{6}\pi^2r_T & -\frac{\pi^2}{6}\frac{\Delta a}{a} & \frac{\pi^2k_\perp^2}{12k_z^2r_T} \\ 0 & 0 & 0 \\ 0 & 0 & 0 \\ 0 & 0 & 0 \\ 0 & 0 & 0 \end{pmatrix}$ | 0 | 0 | $\begin{pmatrix} -\frac{2048r_T}{675} & 0 & \frac{1024k_\perp^2}{675k_z^2r_T} \\ -\frac{2048r_T}{675} & 0 & \frac{1024k_\perp^2}{675k_z^2r_T} \\ -\frac{2048}{675} & 0 & -\frac{2048}{675} \\ 0 & 0 & 0 \\ 0 & 0 & 0 \\ 0 & \frac{4096}{675} & \frac{2048k_\perp^2}{675k_z^2r_T}\frac{\Delta a}{a} \end{pmatrix}$ |

(17)

The following points related to the $\varepsilon^{(p'_x,p'_y,s')\leftrightarrow(p_xp_y,s)}_{(m_xm_y)}$ are worth mentioning:

- A phonon mode with an uneven (even) index $m_i$ induces a strain field with even (uneven) orbital parity along the $i$ axis. In a perfectly square trap, this phonon can thus only efficiently couple polariton modes with the same (different) parity with respect to the $i$ axis. The vanishing $\varepsilon^{(p'_x,p'_y,s')\leftrightarrow(p_xp_y,s)}_{(m_xm_y)}$ terms in Table IV result from combinations of indices, for which this condition is not satisfied.

- As a consequence, phonons with uneven indices can efficiently modulate the energy of the individual states by inducing a strong coupling $g^{(p_xp_y)}_{0,(m_xm_y)}$ (cf. Fig. 13) . This modulation amplitude, which is normally dominated by the $e^c_{zz}$ component, has an important role in the formation of sidebands.

- Phonons with even indices, in contrast, lead to low $g^{(p_xp_y)}_{0,(m_xm_y)}$ couplings. These phonons can, however, induce couplings between levels with different orbital indices with a strength $g^{p_xp_y\leftrightarrow p'_xp'_y}_{\uparrow\uparrow(\uparrow\downarrow),m_xm_y}$ mainly determined by the $e^c_{zz}$ and $e^c_{xz}$ components for spin-conserving and spin-flipping coupling (see below). As an example, a $(m_xm_y) = (2,2)$ phonon can couple both ES modes (such as $(p_xp_y) = (2,1)$ and $(1,2)$) as well as the GS and ES polaritons $(p_xp_y) = (1,1)$ and $(2,2)$ (see Table V). The latter creates a channel for polariton relaxation from the ES to the GS with phonon emission. The coupling is not resonant since the splitting between the polariton states is normally much larger than the phonon energy.

- The uniaxial components $e^c_{xx}$, $e^c_{yy}$, $e^c_{zz}$ do not reduce the symmetry of traps below the tetragonal one and can, therefore, not couple states with different pseudo-spins. This coupling requires shear components. In a perfectly square trap (i.e., with $\Delta a = 0$), the only non-vanishing shear component is the $e^c_{xy}$ one induced by the TA mode (this result also applies for other phonons). The TA phonons are, therefore, the only ones able to generate a $g^{p_xp_y \leftrightarrow p'_xp'_y}_{\uparrow\downarrow,m_xm_y}$ coupling for the efficient interaction of states with different pseudo-spin. A trap distortion can introduce a small component $e^c_{xy} \propto \Delta a/a$ for the LA' mode.

### *3. Strain effects on excitonic states*

The electron states forming the excitons have a simple s-type electronic wavefunction, which is only sensitive to the hydrostatic component of the strain field. The valence band states, in contrast, are superpositions of $p$-like $\langle X\rangle$, $\langle Y\rangle$, and $\langle Z\rangle$ orbitals, which become mixed by the strain. The relevant excitonic levels are those associated with the lowest lying electron-heavy hole (hh) and electron-light hole (lh) transitions in the QWs formed by linear superpositions of the $s$ and $p$-orbitals. We describe these states in a basis of valence band states angular and spin momenta $\langle m_j, s\rangle$ with $j = 3/2$ and $m = 1/2$ given by $(X_{hh\Uparrow}, X_{lh\Downarrow}, X_{lh\Uparrow}, X_{hh\Downarrow})$, which mix with the conduction band electrons to form the excitons.

The effects of the strain on the excitonic states are described by the well-known Pikus and Bir (PB) hamiltonian[10], which in the basis $(X_{hh\Uparrow}, X_{lh\Downarrow}, X_{lh\Uparrow}, X_{hh\Downarrow})$ becomes:

$$\mathbf{H}_{\rm PB} = \begin{pmatrix} \delta E_{\rm hh} & \frac{d}{2}(e^c_{xy} - ide^c_{xz}) & \frac{1}{2}de^c_{yz} & 0 \\ \frac{d}{2}(e^c_{xy} + ide^c_{xz}) & \delta E_{\rm lh} & 0 & \frac{1}{2}de^c_{yz} \\ \frac{1}{2}de^c_{yz} & 0 & \delta E_{\rm lh} & -\frac{d}{2}(e^c_{xy} - ide^c_{xz}) \\ 0 & \frac{1}{2}de^c_{yz} & -\frac{d}{2}(e^c_{xy} + ide^c_{xz}) & \delta E_{\rm hh} \end{pmatrix} \begin{pmatrix} X_{hh\Uparrow} \\ X_{lh\Downarrow} \\ X_{lh\Uparrow} \\ X_{hh\Downarrow} \end{pmatrix}, \quad (18)$$

where

$$\delta E^{(p_xp_y)}_{\rm hh,(m_xm_y)} = \delta E_{\rm hh} = (2e^c_{xx} + e^c_{zz})a_h + (e^c_{zz} - e^c_{xx})b \quad (19)$$

$$\delta E^{(p_xp_y)}_{\rm lh,(m_xm_y)} = \delta E_{\rm lh} = (2e^c_{xx} + e^c_{zz})a_h - (e^c_{zz} - e^c_{xx})b + \Delta_{\rm hh-lh}. \quad (20)$$

In the previous expressions, $\Delta_{\rm hh-lh}$ is the energy splitting between the lh and hh states and the strain components $e^c_{ij}$ are obtained from $\varepsilon^{(p_xp_y)}_{(m_xm_y)} = (e^c_{xx}, e^c_{xx}, e^c_{zz}, e^c_{yz}, e^c_{xz}, e^c_{xy})^T$ in Eq. 15 and displayed for the $(m_xm_y) = (1,1)$ phonon mode in Table V. $a_h$, $b$, and $d$ are, respectively, the hydrostatic, uniaxial, and shear deformation potentials for the GaAs QWs listed in Table III. In the right-most side of the equation, $N_{mec}$ is the number of phonons and $g^{(p_xp_y)}_{0,(\rm m_xm_y)}$ the energy modulation strength by a single phonon. To simplify the notation, we will eliminate the phonon and polariton indices of $\delta E^{(p_xp_y)}_{\rm hh,(m_xm_y)}$ and $\delta E^{(p_xp_y)}_{\rm lh,(m_xm_y)}$ whenever these can be inferred from the context. Note that the uniaxial strain components along the diagonal modulate the energy of the hh and lh states, while the shear ones introduce a reduction in symmetry that couples these states.

By including the exciton-photon coupling with an energy $\Omega_R$ (and $\Omega_R/3$ for the lh states due to the reduced oscillator strength), the polariton Hamiltonian $H_{Pol}$ in the $(X_{hh\Uparrow}, X_{lh\Downarrow}, X_{lh\Uparrow}, X_{hh\Downarrow}, L_L, L_R)$ basis becomes:

$$\mathbf{H}_{\rm Pol}\Psi_{Pol} = \begin{pmatrix} \delta E_{\rm hh} & \frac{1}{2}d(e^c_{xy} - ie^c_{xz}) & \frac{1}{2}de^c_{yz} & 0 & 0 & -\Omega_{\rm R} \\ \frac{1}{2}d(e^c_{xy} + ie^c_{xz}) & \delta E_{\rm lh} & 0 & \frac{1}{2}de^c_{yz} & 0 & -\frac{\Omega_{\rm R}}{3} \\ \frac{1}{2}de^c_{yz} & 0 & \delta E_{\rm lh} & -\frac{1}{2}d(e^c_{xy} - ie^c_{xz}) & \frac{\Omega_{\rm R}}{3} & 0 \\ 0 & \frac{1}{2}de^c_{yz} & -\frac{1}{2}d(e^c_{xy} + ie^c_{xz}) & \delta E_{\rm hh} & \Omega_{\rm R} & 0 \\ 0 & 0 & \frac{\Omega_{\rm R}}{3} & \Omega_{\rm R} & \delta_{\rm CX} & \frac{i\delta_{\rm LT}}{2} \\ -\Omega_{\rm R} & -\frac{\Omega_{\rm R}}{3} & 0 & 0 & -\frac{i\delta_{\rm LT}}{2} & \delta_{\rm CX} \end{pmatrix} \begin{pmatrix} X_{hh\Uparrow} \\ X_{lh\Downarrow} \\ X_{lh\Uparrow} \\ X_{hh\Downarrow} \\ L_L \\ L_R \end{pmatrix}. \quad (21)$$

The upper right $4\times4$ matrix block corresponds to the exciton Hamiltonian of Eq. 18. In the lower left $2\times2$ block, the diagonal term $\delta_{\rm CX}$ is the detuning between the bare photon and exciton states while the off-diagonal terms describe the mixing of the photonic states induced by the TE-TM coupling discussed in the previous section.

We are interested in the lowest energy eigenstates of Eq. 21, as well as on their mixing by the strain field. In order to obtain an analytical expression for these modes, we proceed with a series of simplification steps. We first note that the lh states are blue-shifted relative to the hh states by an energy $\sim \Delta_{\rm hh-lh}$ much larger than the phonon energy.

To address the polariton-phonon coupling, we can then elliminate the lh states from the basis by using perturbartion theory to include their effects on the coupling between the hh and light states. In this way, the dimension of the basis reduces from 6 to 4. Furthermore, we further simplify the expressions by taking into account that the strain components $e_{xz}$ and $e_{yz}$ vanish for the considered polariton modes (cf. Table IV ). The simplified interaction matrix in the reduced basis $(X_{hh\Uparrow}, X_{hh\Downarrow}, L_L, L_R)$ after these transformations reads:

$$\mathbf{H}_c\Psi_c = \begin{pmatrix} \delta E_{\rm hh} & 0 & 0 & -\Omega_{\rm R} + \frac{\Omega_{\rm R} de^c_{xy}}{\Delta_{\rm hh-lh}} \\ 0 & \delta E_{\rm PB,hl} & \Omega_{\rm R} + \frac{\Omega_{\rm R} de^c_{xy}}{6\Delta_{\rm hh-lh}} & 0 \\ 0 & \Omega_{\rm R} + \frac{\Omega_{\rm R} de^c_{xy}}{6\Delta_{\rm hh-lh}} & \delta_{\rm CX} & \frac{i}{2}\delta^{(p_xp_y)}_{\rm LT} \\ -\Omega_{\rm R} + \frac{\Omega_{\rm R} de^c_{xy}}{\Delta_{\rm hh-lh}} & 0 & -\frac{i}{2}\delta^{(p_xp_y)}_{\rm LT} & \delta_{\rm CX} \end{pmatrix} \begin{pmatrix} X_{hh\Uparrow} \\ X_{hh\Downarrow} \\ L_L \\ L_R \end{pmatrix}. \tag{22}$$

Equation 22 can be analytically diagonalized in the absence of the strain field (i.e., for $\varepsilon_{ij} = 0$) and used to determined the unitary transformation matrix in the polariton basis. Furthermore, phonon-mediated interactions between the upper (UP) and lower (LP) polariton states can be neglected since the energy difference between these two states is much larger than the phonon energy. In this way, we arrive to the following interaction Hamiltonian for the phonon-induced coupling between the LP pseudo-spin states $\left(\Psi^{(p_xp_y)}_{\rm LP\Uparrow}, \Psi_{\rm LP\Downarrow}\right)$:

$$\begin{aligned} \mathbf{H}^{(\mathbf{p_x p_y})}_{\rm LP,(m_x m_y)} &= \mathbf{H}^{(\mathbf{p_x p_y})^\Uparrow \leftrightarrow (\mathbf{p_x p_y})^\Downarrow}_{\rm LP,(m_x m_y)} \\ &= \left[-\Omega_R + \frac{1}{2}\delta_{\rm CX} + \frac{1}{3}(p_x^2 + p_y^2)E_{01}\right]\begin{pmatrix} 1 & 0 \\ 0 & 1 \end{pmatrix} \\ &+ \begin{pmatrix} \frac{1}{4}\delta^{(p_xp_y)}_{\rm LT} + \frac{2\delta_{\rm CX} - \delta^{(p_xp_y)}_{\rm LT} + 4\Omega_R}{8\Omega_R}\delta E_{\rm hh} & \frac{\Omega_R}{6\Delta_{\rm hh-lh}} de^c_{\rm xy} \\ \frac{\Omega_R}{6\Delta_{\rm hh-lh}} de^c_{\rm xy} & -\frac{1}{4}\delta^{(p_xp_y)}_{\rm LT} + \frac{2\delta_{\rm CX} + \delta^{(p_xp_y)}_{\rm LT} + 4\Omega_R}{8\Omega_R}\delta E_{\rm hh} \end{pmatrix}. \end{aligned} \tag{23}$$

From the diagonal elements of the previous equation, one obtains the following approximation for the time-averaged splitting between the lowest energy eigenvalues ($E_\Uparrow$ and $E_\Downarrow$):

$$\Delta E^{(p_xp_y)} = E_\Uparrow - E_\Downarrow \approx \delta^{(p_xp_y)}_{\rm LT}\left(\frac{1}{2} - \frac{1}{4}\frac{\delta_{\rm CX}}{\Omega_R},\right) \tag{24}$$

which is valid for $\delta_{\rm CX} \ll \Omega_R$.

### 4. Intra-level coupling

The on-site energy $g^{(p_xp_y)}_{0,\uparrow\uparrow(m_xm_y)}$ as well as the coupling energy between the pseudo-spin states $g^{(p_xp_y)}_{0,\uparrow\downarrow(m_xm_y)}$ induced by a single phonon can be determined directly from the diagonal and non-diagonal elements of Eq. 23, respectively:

$$g^{(p_xp_y)}_{0,\uparrow\uparrow(m_xm_y)} = \left(\frac{1}{2} + \frac{\delta_{\rm CX}}{2\Omega_R}\right)\left[(2e^{\rm c,ZPM}_{xx} + e^{\rm c,ZPM}_{zz})a_h + (e^{\rm c,ZPM}_{zz} - e^{\rm c,ZPM}_{xx})b\right] \tag{25}$$

$$g^{(p_xp_y)}_{0,\uparrow\downarrow(m_xm_y)} = \frac{\Omega_R}{6\Delta_{\rm hh-lh}} de^{\rm c,ZPM}_{xy}. \tag{26}$$

In these expressions, the superscripts ZPMR specify that the strain components are evaluated for a single phonon and we assumed that $|\delta^{(p_xp_y)}_{\rm LT}| \ll \Omega_R$. They justify our previous assertion that the uniaxial strain components defining the diagonal element $\delta E_{hh}$ modulate the energy of the polariton states while the shear component couple their pseudo-spin counterparts $\Psi^{(p_xp_y)}_{\rm LP\Uparrow}$ and $\Psi^{(p_xp_y)}_{\rm LP\Downarrow}$ states. We note that $e^{\rm c,ZPM}_{xy}$ and, thus, the pseudo-spin coupling $g^{(p_xp_y)}_{0,\uparrow\downarrow(m_xm_y)}$, vanishes for the TA' and LA' modes in square traps (i.e., with $\Delta a/a = 0$), as well as for uneven phonons. For the TA mode, in contrast, $g^{(p_xp_y)}_{0,\uparrow\downarrow(m_xm_y)}$ is large and not sensitive to the trap asymmetry.

Equations 25 and 26 apply for the first-order coupling between pseudo-spin states. In the presence of stimulated phonons (e.g., induced by a BAWR), one can also envisage second-order coupling effects between pseudo-spin levels split by twice the phonon energy $\Omega_M$ with an effective coupling strength:

$$G^{(p_xp_y)}_{2,\uparrow\downarrow(m_xm_y)} = \frac{g^{(p_xp_y)}_{0,\uparrow\downarrow(m_xm_y)} g^{(p_xp_y)}_{0,(m_xm_y)}}{\hbar\Omega_M}. \tag{27}$$

Here, the polariton states emits (or absorbs) two phonons, the first in a spin-conserving and the second in a spin-flipping transition. Such a second-order process becomes relevant when $G^{(p_xp_y)}_{2,\uparrow\downarrow(m_xm_y)} \ll g^{(p_xp_y)}_{0,\uparrow\downarrow(m_xm_y)}$.

### 5. *Inter-level coupling*

The framework of the previous section can be readily extended to address the inter-level coupling mechanism between polariton levels with different orbital indices $(p_xp_y)$ and $(p'_x, p'_y)$ mediated by a $(m_xm_y)$ phonon. The interaction between these states is described by the following Hamiltonian:

$$\mathbf{H}^{(\mathbf{p_x p_y})\leftrightarrow(\mathbf{p'_x},\mathbf{p'_y})}_{\mathrm{LP},(\mathrm{m_x m_y})} = \begin{pmatrix} \mathbf{H}^{(\mathbf{p_x p_y})^{\Uparrow}\leftrightarrow(\mathbf{p_x p_y})^{\Downarrow}}_{\mathrm{LP},(\mathrm{m_x m_y})} & \mathbf{H}^{(\mathbf{p_x p_y})^{\Uparrow}\leftrightarrow(\mathbf{p'_x p'_y})^{\Downarrow}}_{\mathrm{LP},(\mathrm{m_x m_y})}\Big|_{\mathbf{mec}} \\ \mathbf{H}^{(\mathbf{p'_x p'_y})^{\Uparrow}\leftrightarrow(\mathbf{p_x p_y})^{\Downarrow}}_{\mathrm{LP},(\mathrm{m_x m_y})}\Big|_{\mathbf{mec}} & \mathbf{H}^{(\mathbf{p'_x p'_y})^{\Uparrow}\leftrightarrow(\mathbf{p'_x p'_y})^{\Downarrow}}_{\mathrm{LP},(\mathrm{m_x m_y})} \end{pmatrix}. \tag{28}$$

The $2\times2$ diagonal blocks of Eq. 28 are defined in Eq. 23. The $2\times2$ off-diagonal ones (i.e., $\mathbf{H}^{(\mathbf{p_x p_y})\leftrightarrow(\mathbf{p'_x},\mathbf{p'_y})}_{\mathrm{LP},(\mathrm{m_x m_y})}\Big|_{\mathbf{mec}}$) are also obtained from Eq. 23 by neglecting the strain-independent terms. This equation defines the following two single-phonon coupling strengths for spin-conserving and a spin-flipping polariton transitions between the $(p_xp_y)$ and $(p'_x, p'_y)$ polariton states:

$$N_{mec} g^{p_xp_y\leftrightarrow p'_xp'_y}_{0,\uparrow\uparrow,m_xm_y} = \mathbf{H}^{(\mathbf{p_x p_y})^{\Uparrow}\leftrightarrow(\mathbf{p_x p_y})^{\Downarrow}}_{\mathrm{LP},(\mathrm{m_x m_y})}|_{1,3} + \mathbf{H}^{(\mathbf{p_x p_y})^{\Uparrow}\leftrightarrow(\mathbf{p_x p_y})^{\Downarrow}}_{\mathrm{LP},(\mathrm{m_x m_y})}|_{2,4} \tag{29}$$

$$N_{mec} g^{p_xp_y\leftrightarrow p'_xp'_y}_{0,\uparrow\downarrow,m_xm_y} = \mathbf{H}^{(\mathbf{p_x p_y})^{\Uparrow}\leftrightarrow(\mathbf{p_x p_y})^{\Downarrow}}_{\mathrm{LP},(\mathrm{m_x m_y})}|_{1,4} + \mathbf{H}^{(\mathbf{p_x p_y})^{\Uparrow}\leftrightarrow(\mathbf{p_x p_y})^{\Downarrow}}_{\mathrm{LP},(\mathrm{m_x m_y})}|_{2,3} \tag{30}$$

## D. Coupling strength and cooperativity

The upper and lower panels of Table V summarized the coupling parameters related to phonon-mediated intra- and inter-level coupling of polariton levels, respectively. The table only includes data for optomechanical transitions with non-vanishing coupling strengths involving the TA', TA, and LA' phonon modes and polaritons with orbital indices $m_i$ and $p_i$ with $i = 1, 2$. The latter were determined using the parameters of Table III for zero photon-exciton detuning, $\delta_{CX} = 0$. The phonon frequencies in the second column were obtained by the diagonalization of Eq. 8. The energy splitting between the states in the $5^{th}$ column depends on the TE-TM splitting (cf. Eq. 24) and is stated as a function of the asymmetry $\frac{\Delta a}{a}$. For the pseudo-spin states, this splitting is typically of only a few $\mu$eV. The $7^{th}$ and $8^{th}$ columns list the level modulation strength $g^{p_xp_y}_{0,m_xm_y}$, which determines the strain-induced modulation amplitude of the level leading to emission sidebands. In agreement with the remarks in Sec. V C 2, the latter is only non-vanishing for the symmetric $(m_xm_y) = 1, 1$ phonons.

### 1. *Coupling and self-oscillations*

Optomechanical self-oscillations require phonon-induced transitions between polariton states spaced by one or two phonon energies, for first and second-order coupling processes, respectively. These transition energies $\Delta E$ between the involved states are listed in the $6^{th}$ column of Table V. Only the $(m_xm_y) = (11)$ phonon can induce intra-level transitions between pseudo-spin states: $\Delta E$ for these transitions is normally small. Inter-level transitions can take place between between GS and ESs as well as between almost degenerated ESs. The former are always of second-order to due the large energy difference ($\Delta E \gg \Omega_{mec}$). The latter can be of first or second-order, depending on the value of $\Delta a/a$. Note, however, that $\Delta E$ can also be significantly changed by a preferential population of one of the pseudo-spin levels via non-linear polariton-polariton interactions.

As a second requirement, self-oscillations pressuposes a sufficiently strong optomechanical coupling to overcome phonon and polariton decoherence. The latter can be quantified by the cooperativity $C^{(p_xp_y)}_{\uparrow\downarrow(\uparrow\downarrow)(m_xm_y)}$, an adimensional parameter giving the optomechanical interaction between phonons and polaritons with finite decoherence rates $\gamma_{\mathrm{pol}}$ and $\Gamma_{\mathrm{mec}}$, respectively. For first-order processes, it is related to the coupling energy $g^{(p_xp_y)}_{(m_xm_y)}$ according to:

TABLE V: Coupling for polariton modes $(p_xp_y)$ and $(p'_x, p'_y)$ induced by $3\lambda$ phonons with orbital indices $(m_xm_y)$, as calculated using the parameters listed in Table III. The phonon frequencies listed in the $2^{nd}$ column were determined using the elastic properties of GaAs. Since the spacer of the MC includes (Al,Ga)As layers, these values sligthly underestimate the measured resonance frequencies. The corresponding frequencies for the $\lambda$ phonons are three times as large. The energy splitting $\Delta E^{(p_xp_y)}$ between the states is give in terms of the trap asymmetry $\frac{\Delta a}{a}$ (cf. Eq. 24). The decay rate for all phonon and polariton modes in the cooperativity calculations were taken to be equal to the one of the ground state polariton and $3\lambda$ mode (see text for details).

INTRA-LEVEL COUPLING

| mode | $f_{\rm BAW}$ (GHz) | $m_xm_y$ | $p_xp_y$ | $\Delta E$ ($\mu eV$) | $g^{p_xp_y}_{0,\uparrow\uparrow m_xm_y}$ (MHz)[a] (MHz)[a] | 1st order $g^{p_xp_y}_{0\uparrow\downarrow,m_xm_y}$ (MHz)[b] | 1st order $1/C^{(max)}_{\uparrow\downarrow}$ | 2nd order $G^{p_xp_y\leftrightarrow p'_xp'_y}_{2\uparrow\downarrow,m_xm_y}$ (MHz)[(c)] | 2nd order $1/C^{(max)}_{2\uparrow\downarrow}$ |
|---|---|---|---|---|---|---|---|---|---|
| TA | 4.559 | 11 | 11 | $-2.38\frac{\Delta a}{a}$ | 0 | 1.15 | 290 | 0. | $\infty$ |
| | 4.559 | 11 | 12 | $-2.11\frac{\Delta a}{a}-1.59$ | 0 | 0.921 | 453 | 0. | $\infty$ |
| | 4.559 | 11 | 22 | $-2.38\frac{\Delta a}{a}$ | 0 | 0.737 | 708 | 0. | $\infty$ |
| LA' | 6.496 | 11 | 11 | $-2.38\frac{\Delta a}{a}$ | 6.68 | $0.0285\frac{\Delta a}{a}$ | $4.73\times10^5(\frac{\Delta a}{a})^{-2}$ | $0.029\times10^{-3}\frac{\Delta a}{a}$ | $1.13\times10^4(\frac{\Delta a}{a})^{-1}$ |
| | 6.496 | 11 | 12 | $-2.11\frac{\Delta a}{a}-1.59$ | 5.35 | $0.0228\frac{\Delta a}{a}$ | $7.39\times10^5(\frac{\Delta a}{a})^{-2}$ | $0.019\times10^{-3}\frac{\Delta a}{a}$ | $1.76\times10^4(\frac{\Delta a}{a})^{-1}$ |

INTER-LEVEL COUPLING

| Mode | $f_{\rm BAW}$ (GHz) | $m_xm_y$ | $p_xp_y$ | $p'_xp'_y$ | $\Delta E$ ($\mu eV$) | $g^{p_xp_y}_{0,m_xm_y}$ (MHz)[a] | $g^{p'_xp'_y}_{0,m_xm_y}$ (MHz)[a] | $g^{p_xp_y\leftrightarrow p'_xp'_y}_{0\uparrow\uparrow,m_xm_y}$ (MHz)[d] | $1/C^{(max)}_{\uparrow\uparrow}$ | $g^{p_xp_y\leftrightarrow p'_xp'_y}_{0\uparrow\downarrow,m_xm_y}$ (MHz) | $1/C^{(max)}_{2,\uparrow\downarrow}$ |
|---|---|---|---|---|---|---|---|---|---|---|---|
| TA' | 4.463 | 11 | 12 | 21 | $2\times10^3\frac{\Delta a}{a}$ | 1.64 | 1.64 | 0.974 | 405 | 0. | $\infty$ |
| | 4.336 | 22 | 21 | 12 | $-2\times10^3\frac{\Delta a}{a}$ | 0. | 0. | 1.27 | 237 | 0. | $\infty$ |
| TA | 4.559 | 11 | 12 | 21 | $2\times10^3\frac{\Delta a}{a}$ | 0. | 0. | $4.26\frac{\Delta a}{a}$ | $21.2(\frac{\Delta a}{a})^{-2}$ | 0. | $\infty$ |
| | 4.673 | 22 | 21 | 12 | $-2\times10^3\frac{\Delta a}{a}$ | 0. | 0. | 0. | $\infty$ | 0.755 | 675 |
| LA | 6.496 | 11 | 12 | 21 | $2\times10^3\frac{\Delta a}{a}$ | 5.35 | 5.34 | 0.0593 | $1.09\times10^5$ | 0. | $\infty$ |
| | 6.664 | 12 | 11 | 12 | $-10^3\,(1+\frac{\Delta a}{a})$ | 0. | 0. | 6.1 | $\infty$ | $0.102^{(f)}$ | $1.28\times10^5$ |
| | 6.815 | 22 | 21 | 12 | $-2\times10^3\frac{\Delta a}{a}$ | 0. | 0. | 5.04 | 15.2 | $0.076\frac{\Delta a}{a}$ | $6.62\times10^4(\frac{\Delta a}{a})^{-1}$ |

(31)

[a] see Eq. 19.
[b] see Eq. 25.
[c] see Eq. 26.
[d] see Eq. 29.
[e] see Eq. 30.
[f] second-order spin-flip process.

$$C^{(p_xp_y)\leftrightarrow(p'_xp'_y)}_{\uparrow\downarrow(m_xm_y)} = N_{pol}C^{(p_xp_y)\leftrightarrow(p'_xp'_y)}_{0,\uparrow\downarrow(m_xm_y)} \quad \text{with} \quad C^{(p_xp_y)\leftrightarrow(p'_xp'_y)}_{0,\uparrow\downarrow(m_xm_y)} = \frac{|g^{(p_xp_y)\leftrightarrow(p'_xp'_y)}_{0,\uparrow\downarrow(m_xm_y)}|^2}{4\gamma_{\rm pol}\Gamma_{\rm mec}} \quad \text{(1st order)}. \tag{32}$$

Here, $N_{pol}$ is the polariton population and $C^{(p_xp_y)\leftrightarrow(p'_xp'_y)}_{0,\uparrow\downarrow(m_xm_y)}$ the single polariton cooperativity. Cooperativities exceeding unity mark the onset of self-oscillations between states split by one (or two, for second-order processes) phonon energies (see, e.g., Ref. [11]). For second-order processes, the cooperativity depends on the populations of the initial and final polariton states[12]. If one assumes both to be equal to $N_{pol}$, then one obtains the following expression for the second-order cooperativity:[12]

$$C^{(p_xp_y)\leftrightarrow(p'_xp'_y)}_{2,\uparrow\downarrow(m_xm_y)} = N_{pol}C^{(p_xp_y)\leftrightarrow(p'_xp'_y)}_{0,\uparrow\downarrow(m_xm_y)} \quad \text{with} \quad C^{(p_xp_y)\leftrightarrow(p'_xp'_y)}_{2,\uparrow\downarrow(m_xm_y)} = \frac{g^{p_xp_y\leftrightarrow p'_xp'_y}_{0,\uparrow\downarrow,m_xm_y}\, g^{(p_xp_y)}_{0,(m_xm_y)}}{4\Omega_{\rm mec}\Gamma_{\rm mec}} \quad \text{(2nd order)}. \tag{33}$$

Note that in contrast to the first-order, the second-order cooperativity does not depend of the polariton decoherence rate.

Columns 9 and 10 (11 and 12) in the upper part of Table V list the coupling energies and associated inverse cooperativities for intra-level coupling of the first (second) order. In the calculations, we assumed that all involved polariton and phonon states have the same decay rate as the ones for the ground state listed in Table III. Following the definition, the inverse cooperativities yield the threshold polariton population for self-oscillations. Remarkably, the $(m_x m_y) = (11)$ TA mode requires less than 1000 polaritons for self-oscillations: this number is significantly smaller than the experimental values for the population in the condensation regime of $10^5$ polaritons.

The spin-flip process involved in the self-oscillations is mediated by the component $e^c_{\rm xy}$ of the TA phonons (see, e.g., Table IV). For LA' modes, this component depends on the trap asymmetry. Note, however, that LA' self-oscillations in traps with with $\Delta a/a < 1$ can also be triggered at polariton populations below the condensation threshold. We remind ourselves that condensation is nevertheless important to ensure long polariton coherences.

Low polariton population thresholds for self-oscillations are also supported by the results for inter-level coupling in the lower part of Table V. Except for the LA' transition $(m_x m_y\, p_x p_y\, p'_x p'_y = 12\,11\,12)$, all relevant couplings are of first-order. Self-oscillations can be induced by both spin-conserving and spin-flitting inter-level transitions mediated by the shear as well as the uniaxial strain components of the phonon mode, respectively (see, e.g., Table IV). As for intra-level coupling, multi-mode self oscillations can in both cases be triggered at polariton populations much less than the condensation threshold. Note, in addition, that spin-conserving processes mediated by the uniaxial components can yield record-high coupling strengths of several MHz, as indicated by the LA' $(m_x m_y, p_x p_y, p'_x p'_y = 12, 21, 12)$ process.

The coupling energy calculations of Table V can be easily extended to $\lambda$ phonons. For the TA mode, the coupling energies do not depend on the phonon wavelength: the conclusions regarding self-oscillations also apply with minor corrections for the $\lambda$ modes. In the case of the LA' mode, the spin-flipping coupling energy reduces by $(1/3)^2$ and the threshold densities increase by approximately one order of magnitude. The latter, however, may be partially compensated by the much longer lifetimes (and higher amplitudes) for the $\lambda$ phonons (cf. Fig. 1).

### E. Quadratic Hamiltonian and strong coupling

#### 1. *Optomechanical parametric oscillator*

We write the Hamiltonian for two polaritonic modes coupled to a phonon as

$$\hat{H} = \hat{H}_0 + H_{\rm int}, \tag{34}$$

$$\hat{H}_0 = \hbar\Omega_M \hat{b}^\dagger \hat{b} + \sum_{j=1,2} \hbar\omega_j \hat{a}_j^\dagger \hat{a}_j, \tag{35}$$

$$H_{\rm int} = H_{{\rm int},1} + H_{{\rm int},2} + H_{{\rm int},h}, \tag{36}$$

$$H_{{\rm int},j} = \hbar g_j \hat{a}_j^\dagger \hat{a}_j \left(\hat{b}^\dagger + \hat{b}\right), \tag{37}$$

$$H_{{\rm int},h} = \hbar g_h \left(\hat{a}_1^\dagger \hat{a}_2 + \hat{a}_2^\dagger \hat{a}_1\right)\left(\hat{b}^\dagger + \hat{b}\right) \tag{38}$$

where $H_{{\rm int},1}$ and $H_{{\rm int},2}$ couple the phonon with each mode separately whereas $H_{{\rm int},h}$ involves a hopping term between the two modes. The strengths of these optomechanical coupling interactions are given by the rates $g_1$, $g_2$ and $g_h$. In these equations $\hat{a}_j^\dagger$ $(\hat{a}_j)$ creates (annihilates) a polariton in the $j$-mode with energy $\hbar\omega_j$ and $\hat{b}^\dagger$ $(\hat{b})$ creates (annihilates) the cavity phonon having energy $\hbar\Omega_M$.

Our goal is to extract the most important mode mixing contribution of the optomechanical coupling $H_{\rm int}$ for the case in which the detuning between the two modes is two phonons, i.e., for $\omega_2 - \omega_1 = 2\Omega_M$. It is easy to see that the energy conserving process for mode mixing at such a detuning requires a second order process involving once $H_{{\rm int},h}$ and once one of the $H_{{\rm int},j}$ mode-conserving interactions: this, due to $j = 1, 2$ and ordering, means four different paths interfering. In order to obtain such a second order contribution we apply the unitary transformation $\tilde{\hat{H}} = \exp(-\hat{S})\hat{H}\exp(\hat{S})$ and $\tilde{\hat{H}} \approx \hat{H}_0 + \frac{1}{2}[\hat{H}_{\rm int}, \hat{S}]$ provided that $[S, \hat{H}_0] = \hat{H}_{\rm int}$. Using the energy eigenbasis of the unperturbed Hamiltonian, $\hat{H}_0|n_1, n_2, n_b\rangle = E_{n_1,n_2,n_b}|n_1, n_2, n_b\rangle$, the above commutator gives the required non-diagonal matrix elements of $\hat{S}$:

$$\langle n'_1, n'_2, n'_b|\hat{S}|n_1, n_2, n_b\rangle = \langle n'_1, n'_2, n'_b|\hat{H}_{\rm int}|n_1, n_2, n_b\rangle/(E_{n_1,n_2,n_b} - E_{n'_1,n'_2,n'_b}). \tag{39}$$

For the detuning condition $\omega_2 - \omega_1 = 2\Omega_M$ the transformed Hamiltonian $\tilde{\hat{H}}$ contains the following energy conserving non-diagonal term

$$\frac{(g_1 - g_2)g_h}{\Omega_M} \hat{a}_2^\dagger \hat{a}_1 \hat{b}^2 + h.c., \tag{40}$$

namely, a polariton transfer from the highest (lowest) energy mode to the lowest (highest) energy mode creates (destroys) two phonons. Thus this synthetic intermode coupling is quadratic in the phonon operators and has an effective coupling rate of $G_2 = (g_1 - g_2)g_h/\Omega_M$. In general $g_1$ is equal to $g_2$, however, different polariton populations of the modes can lead to $g_1 - g_2 \neq 0$ [13]. The above expression for $G_2$, assuming $g_\Delta = g_1 - g_2 = g_{0,\uparrow\uparrow}$, is used at the end of Sec. **Electrically stimulated sidebands** of the main text.

### 2. *Quadratic Hamiltonian and RWA*

On the basis of measurements and arguments in Ref. [14], we assume that there exists an optomechanical coupling between two polaritonic modes (for simplicity in this section we refer to the ground state and the excited state with the subscript $g$ and $e$, respectively) involving a quadratic phonon operator. Such interaction is included in the Hamiltonian

$$\hat{H} = \hbar\Omega_M \hat{b}^\dagger\hat{b} + \sum_{i=g,e} \hbar\omega_i \hat{a}_i^\dagger \hat{a}_i + \hbar G_2 \left(\hat{a}_e^\dagger \hat{a}_g + \hat{a}_g^\dagger \hat{a}_e\right)\left(\hat{b}^\dagger + \hat{b}\right)^2 + \hat{H}_{\text{driving}}. \tag{41}$$

For simplicity we assume a driving Hamiltonian that feeds independently and coherently each of the three modes via the rates $\beta_e$, $\beta_g$, and $\beta_m$ with frequencies $\omega_{e,d}$, $\omega_{g,d}$, and $\Omega_d$, respectively.

$$\hat{H}_{\text{driving}} = \hbar(\beta_m \hat{b} e^{\mathrm{i}\Omega_d t} + \beta_g \hat{a}_g e^{\mathrm{i}\omega_{g,d} t} + \beta_e \hat{a}_e e^{\mathrm{i}\omega_{e,d} t} + h.c.) \tag{42}$$

However, we can also apply the results below to the case in which the polaritonic modes are nonresonantly driven. By moving to the interaction picture of

$$\hat{H}_0 = \hbar\Omega_d \hat{b}^\dagger \hat{b} + \sum_{i=g,e} \hbar\omega_{i,d} \hat{a}_i^\dagger \hat{a}_i, \tag{43}$$

fixing the condition $\omega_{e,d} - \omega_{g,d} = 2\Omega_d$, and discarding the fast oscillating terms we get a RWA Hamiltonian

$$\hat{H} = \hbar\Delta_m \hat{b}^\dagger\hat{b} + \sum_{i=e,g} \hbar\Delta_i \hat{a}_i^\dagger \hat{a}_i + \hbar G_2 \left(\hat{a}_e^\dagger \hat{a}_g \hat{b}^2 + \hat{a}_g^\dagger \hat{a}_e (\hat{b}^\dagger)^2\right) + \hbar(\beta_m \hat{b} + \beta_e \hat{a}_e + \beta_g \hat{a}_g + h.c.) \tag{44}$$

with the detuning variables defined as $\Delta_m = \Omega_M - \Omega_d$ and $\Delta_i = \omega_i - \omega_{i,d}$, with $i = e, g$. Notice that within this RWA picture the choice of $\Delta_m = \Delta_e = \Delta_g = 0$ is the condition that we are interested in since this implies that $\omega_e - \omega_g = 2\Omega_M$, i.e., the detuning between the excited and the ground state is resonant with the energy of two phonons.

### 3. *Strong coupling*

The optomechanical coupling of the Eq.(44) can be linearized considering fluctuations around the strong coherent amplitudes of each mode (arising due to the presence of polaritonic and mechanical driving by a BAWR transducer). First, following a similar procedure used to describe optomechanical strong-coupling for the case of linear optomechanical interaction in Ref. [11], we consider fluctuations in the excited state which is *less* populated than the ground state. For this we replace $\hat{a}_e \to \alpha_e + \delta\hat{a}_e$ and $\hat{a}_g \to \alpha_g$. We take the large numbers $\alpha_{g/e} = \sqrt{n_{g/e}}$ to be real, these strong coherent amplitudes of the polaritonic fields are set by the driving fields. Similarly, we write the mechanical mode operator as $\hat{b} = \alpha_b + \delta\hat{b}$ where $\delta\hat{b}$ is the mechanical fluctuation around the strong coherent amplitude of the mechanical field and take $\alpha_b = \sqrt{n_b}$ real. Indeed, for this detuning condition the system naturally tends to the large $n_b$ regime even in absence of the phonon driving. It can be thought of as a result of polariton induced parametric driving of the phonon field, resembling a single mode squeezing that is limited by the nonlinearity. The amplitude of the mechanical RF driving, $\beta_m$, provides a knob to further increase $n_b$. Neglecting higher order terms and keeping the dominant interaction term for the case of strong mechanical driving, assuming $\sqrt{n_b} > \sqrt{n_e}$, we arrive to the following beam splitter Hamiltonian

$$\hat{H}_{m,e} = \hbar\Delta_m \hat{b}^\dagger\hat{b} + \hbar\Delta_e \delta\hat{a}_e^\dagger \delta\hat{a}_e + \hbar 2G_2\sqrt{n_b n_g}\left(\delta\hat{a}_e^\dagger \delta\hat{b} + \delta\hat{b}^\dagger \delta\hat{a}_e\right), \tag{45}$$

which describes the coherent energy exchange between the fluctuation of the mechanical field and the fluctuation of the polaritonic excited state. Introducing the linewidth of the excited state $\kappa_e$ and the phonon linewidth $\Gamma_m$ the equations of motion are

$$\frac{d}{dt}\begin{pmatrix} \langle\delta a_e\rangle \\ \langle\delta b\rangle \end{pmatrix} = -\mathrm{i}\begin{pmatrix} \Delta_e - \mathrm{i}\frac{\kappa_e}{2} & 2\sqrt{n_b n_g}G_2 \\ 2\sqrt{n_b n_g}G_2 & \Delta_m - \mathrm{i}\frac{\Gamma_m}{2} \end{pmatrix} \cdot \begin{pmatrix} \langle\delta a_e\rangle \\ \langle\delta b\rangle \end{pmatrix}. \tag{46}$$

Defining $g = 2\sqrt{n_b n_g} G_2$, $\delta = \Delta_e - \Delta_m$ and $\gamma = \Delta_e + \Delta_m$ the eigenfrequencies become

$$\omega_\pm = \frac{\gamma}{2} - \mathrm{i}\frac{\kappa_e + \Gamma_m}{4} \pm \sqrt{g^2 + \left(\frac{\delta + \mathrm{i}(\Gamma_m - \kappa_e)/2}{2}\right)^2} \tag{47}$$

For $\Gamma_m \ll \kappa_e$ and at resonance, $\Delta_e = \Delta_g = \Delta_m = 0$, the expression simplifies to

$$\omega_\pm = -\mathrm{i}\frac{\kappa_e}{4} \pm \sqrt{g^2 - \frac{\kappa_e^2}{4^2}} \tag{48}$$

and one can see that strong coupling arises when the effective interaction,

$$g = 2\sqrt{n_b n_2} G_2 > \kappa_e/4. \tag{49}$$

The imaginary part of $\omega_\pm$ being $\kappa_e/4$ implies that the peak linewidth is $\kappa_e/2$, i.e., a 50% reduction. This is due to the fact that these excitations are half mechanical and the phonon lifetime, being much larger than $\kappa_e^{-1}$, virtually does not contribute to the linewidth. Importantly, the compounded mechanical and optical enhancements of the effective coupling constant relaxes the number of phonons and polaritons required for reaching the strong coupling regime. The above procedure is similar to that describing strong coupling in the case of a linear optomechanical interaction as presented in Ref. [11].

We proceed further by noting that the optomechanical interaction of the RWA Hamiltonian can be written as $\hbar G_2(\hat{\Psi}^\dagger \hat{b}^2 + \hat{\Psi}(\hat{b}^\dagger)^2)$ with the operator $\hat{\Psi} = \hat{a}_g^\dagger \hat{a}_e$. It is easy to see that in absence of such interaction this polaritonic operator, decays with $(\kappa_e + \kappa_g)/2$ with $\kappa_g$ the decay rate of the ground state. Since $[\hat{\Psi}, \hat{\Psi}^\dagger] = \hat{a}_g^\dagger \hat{a}_g - \hat{a}_e^\dagger \hat{a}_e + 1$, it is convenient to define $\tilde{\hat{\Psi}} = \hat{\Psi}/\sqrt{n_g - n_e + 1}$, so that in the high occupation limit $[\tilde{\hat{\Psi}}, \tilde{\hat{\Psi}}^\dagger] = (\hat{a}_g^\dagger \hat{a}_g - \hat{a}_e^\dagger \hat{a}_e + 1)/(n_g - n_e + 1) = 1$, i.e., this polaritonic excitation fulfills the bosonic commutation relation. Introducing this operator and the expansion $\hat{b} = \sqrt{n_b} + \hat{\delta b}$ in Eq.(44) we obtain a beam-splitter-like Hamiltonian

$$\frac{d}{dt}\begin{pmatrix} \langle \delta\tilde{\Psi} \rangle \\ \langle \delta b \rangle \end{pmatrix} = -\mathrm{i}\begin{pmatrix} \Delta_e - \Delta_g - \mathrm{i}\frac{\kappa_e + \kappa_g}{2} & 2\sqrt{n_b(n_g - n_e + 1)} G_2 \\ 2\sqrt{n_b(n_g - n_e + 1)} G_2 & \Delta_m - \mathrm{i}\frac{\Gamma_m}{2} \end{pmatrix} \cdot \begin{pmatrix} \langle \delta\tilde{\Psi} \rangle \\ \langle \delta b \rangle \end{pmatrix}. \tag{50}$$

For $\Gamma_m \ll \kappa_e + \kappa_g$ and at resonance, $\Delta_e = \Delta_g = \Delta_m = 0$, the eigenenergies simplifies to

$$\omega_\pm = -\mathrm{i}\frac{\kappa_e + \kappa_g}{4} \pm \sqrt{\tilde{g}^2 - \frac{(\kappa_e + \kappa_g)^2}{4^2}} \tag{51}$$

For $n_g \gg n_e$, $\tilde{g} = 2\sqrt{n_b n_g} G_2$ and one recovers the same optomechanical effective coupling as in Eq.49. For $\tilde{g} > (\kappa_e + \kappa_g)/4$ these solutions have linewidth $(\kappa_e + \kappa_g)/4$, i.e., half the original linewidth. Since the excited state linewidth was shown to halve when $2\sqrt{n_b n_g} G_2 \gg \kappa_e/4$ the halving of the sum of the linewidths indicates that the effective linewidth of the ground state also halves. We note that the operator $\tilde{\hat{\Psi}}$ can be interpreted as describing a quasi-particle that represents the beating between states $g$ and $e$, and which is resonantly coupled with the mechanics through the creation of two phonons.